\documentclass[structabstract]{aa}  
\usepackage{graphicx}
\usepackage{float}
\usepackage{indentfirst}
\usepackage{lscape}
\usepackage{longtable}
\usepackage{txfonts}
\begin{document}

    \title{Surface chemistry in photodissociation regions}

   \author{G. B. Esplugues
          \inst{1},
          S. Cazaux  
          \inst{1},  
          R. Meijerink
          \inst{2},
          M. Spaans
          \inst{1},
          \and
          P. Caselli
          \inst{3}
         }

   \institute{Kapteyn Astronomical Institute, University of Groningen, P.O. Box 800, NL 9700 AV Groningen, The Netherlands\\
              \email{esplugues@astro.rug.nl}
        \and
            Leiden Observatory, Leiden University, P.O. Box 9513, NL 2300 RA Leiden, The Netherlands
        \and
Max Planck Institute for Extraterrestrial Physics, Giessenbachstrasse 1, D-85748 Garching, Germany.
             }

   \date{Received ; accepted }

 
  \abstract   
   {The presence of dust can strongly affect the chemical composition of the interstellar medium. We model the chemistry in photodissociation regions (PDRs) using both gas-phase and dust-phase chemical reactions. 
   } 
   {Our aim is to determine the chemical compositions of the interstellar medium (gas/dust/ice) in regions with distinct (molecular) gas densities that are exposed to radiation fields with different intensities.}
   {We have significantly improved the Meijerink PDR code by including 3050 new gas-phase chemical reactions and also by implementing surface chemistry. In particular, we have included 117 chemical reactions occurring on grain surfaces covering different processes, such as adsorption, thermal desorption, chemical desorption, two-body reactions, photo processes, and cosmic-ray processes on dust grains.} 
   {We obtain abundances for different gas and solid species as a function of visual extinction, depending on the density and radiation field. We also analyse the rates of the formation of CO$_2$ and H$_2$O ices in different environments. In addition, we study how chemistry is affected by the presence/absence of ice mantles (bare dust or icy dust) and the impact of considering different desorption probabilities.}
   {The type of substrate (bare dust or icy dust) and the probability of desorption can significantly alter the chemistry occurring on grain surfaces, leading to differences of several orders of magnitude in the abundances of gas-phase species, such as CO, H$_2$CO, and CH$_3$OH. The type of substrate, together with the density and intensity of the radiation field, also determine the threshold extinction to form ices of CO$_2$ and H$_2$O. We also conclude that H$_2$CO and CH$_3$OH are mainly released into the gas phase of low, far-ultraviolet illuminated PDRs through chemical desorption upon two-body surface reactions, rather than through photodesorption.}

   \titlerunning{Surface chemistry in photodissociation regions}
   \authorrunning{G. B. Esplugues et al.}
   \maketitle
%

\section{Introduction}

Photodissociation regions (PDRs) consist of predominantly neutral gas and dust illuminated by far-ultraviolet (FUV) radiation (6$<$h$\nu$$<$13.6 eV). Dense PDRs are found in the vicinity of star-forming regions, since FUV photons usually arise from massive stars creating HII regions. Ultraviolet (UV) photons impinge on clouds of gas and dust and play an important role in the heating and chemistry of these irradiated regions; while the gas is heated to relatively high temperatures, the ionisation and, mainly, the photodissociation of different species are produced as the UV radiation penetrates into the region (Hollenbach et al. 1997). This, therefore, produces significant differences in the chemical composition of the cloud depending on the visual extinction.

To properly model the chemical composition of PDRs, it is key to take the role of dust in the chemistry of these regions into account. In environments powered by high radiation fields, the dust grains are mainly bare, since no ice mantles can form on their surfaces owing to radiation (Meijerink et al. 2012). Dust grains, when not covered by ice, provide an ideal place for chemical reactions to occur that directly enrich the gas phase. For higher extinctions, however, dust grains are mainly covered by ice mantles, which usually sublimate into the gas phase by star formation activities (Viti et al. 2004), enhancing the chemical composition of the gas phase as well.

The first numerical models of PDRs were developed by Hollenbach et al. (1971), Glassgold \& Langer (1975), and Black \& Dalgarno (1977), who assume a steady state to simulate the transitions from H to H$_2$ and from C$^+$ to CO. Later, new models, such as those from van Dishoeck \& Black (1988), Le Bourlot et al. (1993), St\"orzer et al. (1996), and Spaans (1996) were developed to focus on the chemical and thermal structure of clouds subject to an incident flux of FUV radiation. More recently, models considering time-dependent chemical networks (e.g. Bertoldi \& Draine 1996), turbulence (e.g. R\"ollig et al. 2002; Bell et al. 2005) and 3D (Bisbas et al. 2012) have also been developed.   
All these models are mainly based on gas-phase chemistry. In fact, only a few PDR codes consider reactions on dust surfaces (apart from H$_2$ formation). These include the {\it{Meudon}} PDR code from Le Petit et al. (2006) and the PDR code from Hollenbach et al. (2009).

Many recent observations (e.g. Bern\'e et al. 2007; Sellgren et al. 2010; Guzm\'an et al. 2013; Cuadrado et al. 2015) show the chemical richness of dense PDRs. This richness makes it evident that there is a need to consider a larger number of solid species in the current codes
to understand the role of dust in the origin of the chemical complexity of PDRs, since surface chemical reactions can dramatically alter the gas-phase composition of these regions. It is, therefore, crucial to include a detailed treatment of the chemistry occurring on dust grains to really understand the link between dust chemistry and gas chemistry. This treatment should not only consider the most relevant physical processes for grains at large visual extinctions, but also study the presence or absence of ice mantles on their surfaces.

In this paper, we have significantly improved the chemical network of the {\it{Meijerink}} code (Meijerink \& Spaans 2005), by considering 3050 new gas-phase chemical reactions and, mainly, by implementing dust grain chemistry to determine the different compositions in PDRs (gas/dust/ice). These chemical treatments include adsorption, thermal and chemical desorption, two-body reactions, photo processes, and cosmic-ray processes on dust grains, as well as the presence of some complex organic solid species, such as H$_3$CO and CH$_3$OH. 
In Sect. \ref{Numerical_mode}, we explain the new chemical treatment, especially for dust grains depending on the presence or absence of ice mantles, and we present the results in Sect. \ref{Results}. In Sect. \ref{Discussion}, we analyse the formation routes for ices of CO$_2$ and H$_2$O depending on the environmental conditions. We also study the impact of the presence of dust grains and of their type of substrate (bare or icy) on the chemical composition of regions powered by UV photons. In addition, we compare our results with observations from the Horsehead PDR and the Orion Bar PDR in Sect. \ref{Comparison_observations}. Finally, we summarise the main conclusions in Sect. \ref{conclusions}.

\section{The numerical code}
\label{Numerical_mode}

\subsection{Gas chemistry}
\label{Gas_chemistry}

We consider in our steady-state PDR code 7503 gas-phase chemical reactions from the Kinetic Database for Astrochemistry (KIDA; Wakelam et al. 2015)\footnote{http://kida.obs.u-bordeaux1.fr}, including bimolecular reactions (A + B $\rightarrow$ C + D), charge-exchange reactions (A$^{+}$ + B $\rightarrow$ A + B$^{+}$), radiative associations (A + B $\rightarrow$ AB + photon), associative detachment (A$^{-}$ + B $\rightarrow$ AB + e$^{-}$), dissociative recombination (A$^{+}$ + e$^{-}$ $\rightarrow$ C + D), neutralisation reactions (A$^{+}$ + B$^{-}$ $\rightarrow$ A + B), ion-neutral reactions (A$^{+}$ + B $\rightarrow$ C$^{+}$ + D), ionisation or dissociation of neutral species by UV photons, and ionisation or dissociation of species by direct collision with cosmic-ray particles or by secondary UV photons following H$_2$ excitation. The initial gas-phase abundances ($A$$_{\mathrm{i}}$) for the different elements that we consider (Jenkins 2004, Asplund et al. 2005, and Neufeld \& Wolfire 2009) are listed in Table \ref{table:abundances}.

The thermal balance of PDRs is determined by different heating and cooling processes. As heating mechanisms, we consider the photoelectric effect on grains, carbon ionisation heating, H$_2$ photodissociation heating by UV photons, H$_2$ collisional de-excitation heating, gas-grain collisional heating, gas-grain viscous heating\footnote{Radiation pressure accelerates grains relative to the gas and the resulting drag contributes to viscous heating to gas.}, and cosmic-ray heating. As cooling mechanisms, we consider fine-structure line cooling (being [CII] at 158 $\mu$m and [OI] at 63 $\mu$m and at 146 $\mu$m the most prominent cooling lines), metastable-line cooling (including lines of C, C$^+$, Si, Si$^+$, O, O$^+$, S, S$^+$, Fe, and Fe$^+$), recombination cooling, and molecular cooling by H$_2$, CO, and H$_2$O (see Meijerink \& Spaans 2005 for more details of each process).

\begin{table}[h!]
\caption{Abundances with respect to number density of H nuclei. 
}             
\centering 
\begin{tabular}{l l l l l}     
\hline\hline       
Species &  $A$$_{\mathrm{i}}$ (gas) & \vline  & Species &  $A$$_{\mathrm{i}}$ (gas)                       \\ 
\hline 
H   & 1.0                   & \vline   &  Cl  & 1.8$\times$10$^{-7}$       \\
C   & 2.5$\times$10$^{-4}$  & \vline   &  Fe  & 2.0$\times$10$^{-7}$    \\
N   & 7.2$\times$10$^{-5}$  & \vline   &  P   & 3.9$\times$10$^{-8}$\\
O   & 4.7$\times$10$^{-4}$  & \vline   &  Na  & 5.9$\times$10$^{-7}$  \\
Si  & 1.7$\times$10$^{-6}$  & \vline   &  Mg  & 3.4$\times$10$^{-6}$   \\
S   & 6.9$\times$10$^{-6}$  & \vline   &  F   & 1.8$\times$10$^{-8}$  \\
He  & 8.5$\times$10$^{-2}$  & \vline   &      &   \\
\hline 
\label{table:abundances}                 
\end{tabular}
\end{table}

\subsection{Dust chemistry}
\label{Dust_chemistry}

While the  gas-phase chemical network that we consider in our code is taken from KIDA 2014, we derived the surface chemical network from laboratory experiments (e.g. Dulieu et al. 2013; Minissale et al. 2015, 2016). The solid species that we consider in the code to model the dust chemistry are listed in Table \ref{table:solid_species}. 
We included 117 chemical reactions occurring on grain surfaces covering different processes (adsorption, thermal and non-thermal desorption, two-body reactions, photo processes, and cosmic-ray processes). These processes are detailed in the following subsections.

\begin{table}[h!]
\caption{Solid species.}             
\centering 
\begin{tabular}{l l l l l }     
\hline\hline       
Species & \vline  & Species & \vline  & Species         \\ 
\hline 
H                   & \vline & O$_3$      &   \vline & HCO         \\
H$_{\mathrm{c}}$    & \vline & HO$_2$     &   \vline & H$_2$CO     \\
O                   & \vline & H$_2$O     &   \vline & H$_3$CO   \\
H$_2$               & \vline & H$_2$O$_2$ &   \vline & CH$_3$OH  \\
OH                  & \vline & CO         &   \vline & N            \\
O$_2$               & \vline & CO$_2$     &   \vline & N$_2$      \\
\hline 
\label{table:solid_species}                 
\end{tabular}
\end{table}

\subsubsection{Adsorption onto dust grains}
\label{Adsorption}

Gas-phase species can be adsorbed on the dust grain surface. This adsorption is determined by the dynamics of the accretion processes acting on the dust in a determined region. In dense regions of the interstellar medium (ISM), accretion is favoured by the increased collision rates between gas-phase species and grains.  
Accretion efficiently depletes grains with a radii lower than 0.001$\mu$m on a timescale of $\lesssim$10 Myr in solar-metallicity molecular clouds with densities $n$$\sim$10$^3$ cm$^{-3}$ (Hirashita 2000, 2012).
The adsorption rate, $R$$_{\mathrm{ad}}$ (cm$^{-3}$s$^{-1}$), of the species $i$ is determined by the total cross section from dust\footnote{The total cross section is obtained by integrating over the grain-size distribution.}, $\sigma$$_{\mathrm{d}}$$n$$_{\mathrm{d}}$ (cm$^{-1}$), the thermal velocity of the species $i$ (v$_{\mathrm{th}_i}$), and the sticking coefficient $S$($T$$_{\mathrm{g}}$, $T$$_{\mathrm{d}}$). The adsorption rate can be written as

\textbf{
\begin{equation}
R_{\mathrm{ad}} = n_{{i}} \sigma_{\mathrm{d}}n_{\mathrm{d}} {\mathrm{v}}_{\mathrm{th}_i}S(T_{\mathrm{g}}, T_{\mathrm{d}}) = n_{{i}} k_{ads} S(T_{\mathrm{g}}, T_{\mathrm{d}}),  
\end{equation}
}

\noindent where $k$$_{ads}$ is the adsorption rate coefficient, $n$$_{{i}}$ is the number density of adsorbing species $i$, and the thermal velocity is written as

\textbf{
\begin{equation}
{\mathrm{v}}_{\mathrm{th}_i}=  \displaystyle{ \sqrt {\frac{8k_{\mathrm{B}}T_{\mathrm{g}}}{\pi m_{i}}}}, 
\end{equation}
}

\noindent where $m$$_{i}$ is the mass in grams of species $i$ and $k$$_{\mathrm{B}}$ the Boltzmann constant. The sticking coefficient for all the species is given by

\textbf{
\begin{equation}
S(T_{\mathrm{g}}, T_{\mathrm{d}}) =  \left(1 + 0.4  \displaystyle{ \sqrt {\frac{T_{\mathrm{g}}+T_{\mathrm{d}}}{100}}} + 0.2 \frac{T_{\mathrm{g}}}{100} + 0.08 \left(\displaystyle{ \frac{T_{\mathrm{g}}}{100}}\right)^{2}\right)^{-1}
\end{equation}
}

\noindent (Hollenbach \& McKee 1979), where $T$$_{\mathrm{g}}$ and $T$$_{\mathrm{d}}$ are the gas and dust temperatures, respectively. In this paper, we use a mean grain cross section $\sigma$$_{\mathrm{MRN}}$=$\langle$$\sigma$$_{\mathrm{d}}$$n$$_{\mathrm{d}}$/$n$$_{\mathrm{H}}$$\rangle$$_{\mathrm{MRN}}$=10$^{-21}$ cm$^{2}$ computed assuming a MRN grain size distribution (Mathis et al. 1977), with grain radius extending from $\sim$50 \AA $\ $ to $\sim$0.25 $\mu$m.
In Table \ref{table:adsorption_reactions} (Appendix), we list the adsorption reactions considered.

\subsubsection{Thermal desorption}
\label{Thermal_desorption}

Once species are depleted on dust grains, they can evaporate back into the gas. This process depends on the dust temperature and on whether the surfaces of the grains are covered by ice (icy grains) or not (bare grains), since the binding energies in both cases are different. We calculate the fraction of the dust that is bare, $f$$_{\mathrm{bare}}$, and icy, $f$$_{\mathrm{ice}}$, taking into account the density of sites, $n$$_{\mathrm{d}}$$n$$_{\mathrm{sites}}$ (cm$^{-3}$), where species are locked on the dust. The density of sites can be written as

\textbf{
\label{equation:ndnsites} 
\begin{equation}
n_{\mathrm{d}}n_{\mathrm{sites}} = n_{\mathrm{d}} \displaystyle{ \frac{4 \pi r_{\mathrm{d}}^2}{(a_{pp})^2}} = n_{\mathrm{d}} \displaystyle{ \frac{4\sigma_{\mathrm{d}}}{(a_{pp})^2}} \simeq 4.44 \times 10^{-6}n_{\mathrm{H}},  
\end{equation}
}

\noindent where $r$ is the radius of dust and $a$$_{pp}$ is the distance between two sites that we assume to be 3$\AA$. This is the typical size between sites that should be considered to obtain a typical site density of $\sim$10$^{15}$ sites/cm$^{2}$. 
A full monolayer is formed when all the possible sites on a grain surface are occupied by an atom or molecule. It means that for solid species, as abundances are higher than 4.44$\times$10$^{-6}$, more than one monolayer is reached. When the grain surface is covered by less than one monolayer, we calculate $f$$_{\mathrm{ice}}$ as

\textbf{
\begin{equation}
f_{\mathrm{ice}}= \displaystyle{ \frac{n_{J(H_{2}O)}}{n_{\mathrm{d}}n_{\mathrm{sites}}}}, 
\end{equation}
}

\noindent where $n$$_{J(H_{2}O)}$ is the number density of solid H$_2$O. If the grain is covered by more than one layer of water ice (i.e. when $n$$_{J(H_{2}O)}$$>$$n$$_{\mathrm{d}}$$n$$_{\mathrm{sites}}$), $f$$_{\mathrm{ice}}$=1. The expression for $f$$_{\mathrm{bare}}$ can be obtained by

\textbf{
\begin{equation}
f_{\mathrm{bare}}= 1-f_{\mathrm{ice}}.
\end{equation}
}

\noindent As previously mentioned, the fraction of bare or icy dust has important consequences for the binding energies of the species. We, therefore, consider two types of binding energies: on bare dust ($E$$_{\mathrm{b}}$) and on icy surfaces ($E$$_{\mathrm{i}}$). The desorption rate, $R$$_{\mathrm{des}}$ (cm$^{-3}$s$^{-1}$), can be written as

\textbf{
\begin{equation}
R_{\mathrm{des}}= \nu_0 n_{{i}} \left[f_{\mathrm{bare}} exp\left(\displaystyle{ \frac{-E_{\mathrm{b}}}{T_{\mathrm{d}}}}\right) + f_{\mathrm{ice}} exp\left(\displaystyle{ \frac{-E_{\mathrm{i}}}{T_{\mathrm{d}}}}\right)\right].
\end{equation}
}

\noindent $\nu$$_0$ is the oscillation frequency that is determined by

\textbf{
\begin{equation}
\nu_0= \displaystyle{ \sqrt {\frac{2 N_{\mathrm{s}} E}{ \pi^{2}m}}}\,, 
\end{equation}
}

\noindent where $N$$_{\mathrm{s}}$ is the surface number density of sites on the grain, $m$￼ is the mass of the species, and $E$ is the energy of the site where the species is bound. $\nu$$_0$ is typically $\sim$10$^{12}$ s$^{-1}$ for physisorbed species. 
In Tables \ref{table:desorption_reactions} and \ref{table:binding_energies}, we list the desorption reactions considered and the binding energies for each species, respectively.

\subsubsection{Two-body reactions on grain surfaces}
\label{Reactions_on_grains}

Once gas species are adsorbed onto dust grains, they can also move around the grain surfaces. There are two different types of interaction between the species and the surface: physisorption and chemisorption. The physisorption is a weak interaction due to Van der Waals forces between the adsorbed atom and the surface (dipole-dipole interaction). The typical depth of wells associated with physisorption are of the order of 0.01-0.2eV (Vidali et al. 1991). The chemisorption is adsorption in which the forces involved are similar to valence forces, and the interaction potential depends not only on the distance from the surface, but also on the position on the surface. The typical binding energies for chemisorption are of the order of $\sim$1 eV (Barlow \& Silk 1976; Zangwill 1988).

The surface of a dust grain is mainly irregular, with the presence of peaks and valleys. The valleys represent the physisorbed (highest) and chemisorbed (deepest) wells and they are separated by saddle points (diffusion barriers). The mobility of adsorbed species is associated with transfer across these barriers, which can occur through thermal diffusion.
Dulieu et al. (2013) and Collings et al. (2003) find experimentally that diffusion occurs with a barrier of 67$\%$ and 40$\%$, resepctively, of the binding energy. Theoretical results show percentages of 90$\%$ (Barzel $\&$ Biham 2007) and 30$\%$ (Karssemeijer $\&$ Cuppen 2014). In this paper, we assume that diffusion occurs with a barrier of 2/3 of the binding energy. The adsorbed species $i$ and $j$ have different probabilities of mobility depending on whether the grain is bare ($P$$_{\mathrm{bare}}$) or icy ($P$$_{\mathrm{ice}}$) as follows:


\begin{equation}
P_{\mathrm{bare}}= f_{\mathrm{bare}} \left[exp\left(\displaystyle{ \frac{-2 E_{\mathrm{b}}(i)}{3T_{\mathrm{d}}}}\right) + exp\left(\displaystyle{ \frac{-2 E_{\mathrm{b}}(j)}{3 T_{\mathrm{d}}}}\right)\right],  
\end{equation}

\noindent and

 \begin{equation}
P_{\mathrm{ice}}= f_{\mathrm{ice}} \left[exp\left(\displaystyle{ \frac{-2 E_{\mathrm{i}}(i)}{3T_{\mathrm{d}}}}\right) + exp\left(\displaystyle{ \frac{-2 E_{\mathrm{i}}(j)}{3 T_{\mathrm{d}}}}\right)\right].  
\end{equation}

\noindent The two-body reaction rate, $R$$_{\mathrm{2body}}$ (cm$^{-3}$s$^{-1}$), is therefore given by

\begin{equation}
R_{\mathrm{2body}}= \nu_0 \left(\displaystyle{ \frac{n_{\mathrm{i}}n_{\mathrm{j}}}{n_{\mathrm{d}}n_{\mathrm{sites}}}}\right) k_{\mathrm{2body}} P_{\mathrm{react}},  
\end{equation}

\noindent where $\nu$$_0$ is the oscillation factor, $k$$_{\mathrm{2body}}$ is the two-body rate coefficient

\begin{equation}
k_{\mathrm{2body}}=  P_{\mathrm{bare}} \delta_{\mathrm{bare}} + P_{\mathrm{ice}} \delta_{\mathrm{ice}}, 
\end{equation}

\noindent and $\delta$$_{\mathrm{bare}}$ and $\delta$$_{\mathrm{ice}}$ are the theoretical probabilities of desorption upon reaction for bare and icy\footnote{$\delta$$_{\mathrm{ice}}$ coefficients are considered to be 10$\%$ of $\delta$$_{\mathrm{bare}}$.} substrates, respectively (Minissale et al. 2016). 
To determine the probability, $P$$_{\mathrm{react}}$, of overcoming a reaction barrier with energy $\epsilon$ (K), we need to consider thermal diffusion, $P$$_{\mathrm{react}}$(therm), and tunneling, $P$$_{\mathrm{react}}$(tunn) as follows:


\begin{equation}
P_{\mathrm{react}}(therm)=  exp\left(\displaystyle{ \frac{-\epsilon}{T_{\mathrm{d}}}}\right),  
\end{equation}

\noindent and

\begin{equation}
P_{\mathrm{react}}(tunn)= exp\left(-a\displaystyle{ \sqrt {\frac{2m_{red}k_{\mathrm{B}}\epsilon}{\hbar^{2}}}}\right), 
\end{equation}

\noindent where $a$ is the width of the barrier of 1 $\AA$ (for a square barrier; Tielens \& Hagen 1982; Hasegawa et al. 1992), $\hbar$ is the Planck constant divided by 2$\pi$, and $m$$_{\mathrm{red}}$ is the reduced mass of the reaction between two species $i$ and $j$, $m$$_{\mathrm{red}}$=($m$$_{\mathrm{i}}$$\times$$m$$_{\mathrm{j}}$)/($m$$_{\mathrm{i}}$+$m$$_{\mathrm{j}}$). The probability $P$$_{\mathrm{react}}$ can be written as

\begin{equation}
P_{\mathrm{react}}= \displaystyle{ \frac{P_{\mathrm{max}}}{P_{\mathrm{max}} + P_{\mathrm{bare}} + P_{\mathrm{ice}}}}, 
\end{equation}

\noindent with $P$$_{\mathrm{max}}$ the maximum between $P$$_{\mathrm{react}}$(tunn) and $P$$_{\mathrm{react}}$(therm). Table \ref{table:two-body-reactions} lists the considered reactions occurring on grain surfaces and the parameters $\delta$$_{\mathrm{bare}}$ and $\delta$$_{\mathrm{ice}}$.
We also consider reactions on dust grains between physisorbed and chemisorbed hydrogen to form molecular hydrogen. See Appendix \ref{Chemisorption} for more details.

\subsubsection{Photo processes on dust grains}
\label{Photo-processes_on_grains}

In the outskirts of molecular clouds and in the vicinity of high-mass stars, far-ultraviolet photons can dominate the chemistry. These photons usually impinge on dust grains located in the surroundings of the forming stars and can lead to the photodesorption and/or photodissociation of the species adsorbed onto dust grains. Since recent results (Mu\~noz-Caro et al. 2010; Fayolle et al. 2011; Chen et al. 2014) indicate that photons adsorbed deeper than a few monolayers have no effect on the photodesorption because they cannot transfer their energy to the uppermost monolayers, in our code we consider that incident photons can only interact the first two layers of ice and produce photodesorption. 

Since photoreactions scale linearly with the local radiation flux (erg cm$^{-2}$ s$^{-1}$) and the radiation field strength is a function of extinction ($\xi$$_{{i}}$A$_V$), where $\xi$$_{{i}}$ is the extinction factor for the relevant species, the photo-process reaction rate, $R$$_{\mathrm{photo}}$ (cm$^{-3}$ s$^{-1}$), can be written as

\begin{equation}
k_{\mathrm{photo}}= \alpha_{{i}} e^{-\xi_{{i}}A_{V}}, 
\end{equation}

\begin{equation}
R_{\mathrm{photo}}= n_{{i}}f_{ss}k_{photo}F_{UV}. 
\end{equation}

\noindent The parameter $n$$_{\mathrm{i}}$ is the number density of the photodissociated species, $k$$_{\mathrm{photo}}$ is the photo-process rate coefficient, $\alpha$$_{{i}}$ is the unattenuated rate coefficient, $f$$_{ss}$ is the self-shielding factor, and $F$$_{\mathrm{UV}}$ is the UV flux in units of 1.71$G$$_0$ ($G$$_0$=1 gives $F$$_{\mathrm{UV}}$=0.58). The factor 1.71 arises from the conversion of the often used Draine field (Draine 1978) to the Habing field for the far-ultraviolet intensity.
We consider self-shielding for H$_2$ and CO molecules (van Dishoeck \& Black 1998).
For the rest of the species, we assumed $f$$_{ss}$=1. Table \ref{table:photo-reactions} lists the reactions occurred on grain surfaces due to photon impacts and the parameters $\alpha$$_{{i}}$ and $\xi$$_{{i}}$. In particular, we consider in our code direct photodesorption of CO, H$_2$O, and H$_2$CO. We do not include direct photodesorption of CH$_3$OH, given that recent laboratory results (Bertin et al. 2016) conclude that this mechanism is not very efficient to release methanol into the gas phase.

\subsubsection{Cosmic-ray processes on dust grains}
\label{UV_cosmic-rays_on_grains}

Cosmic rays impacting on dust grains can provide non-thermal energy to desorb molecules frozen on grain surfaces. In addition, given that cosmic rays have larger penetrating power than UV photons and X-rays, they can have a greater impact on the chemistry in well-shielded regions. In particular, cosmic rays can heat dust grains partially or completely leading to explosive desorption (d'Hendecourt et al. 1982; L\'eger et al. 1985; Ivlev et al. 2015). The grain temperature increase due to cosmic rays depends on their flux, the projected area of the grain, and the energy lost by a cosmic ray as it passes through the grain. Since some electrons excited by cosmic rays have large energies and escape from the grains leading to a reduction of the effective heating, cosmic rays need to pass through a sufficiently long path in the grains to deposit enough energy to produce desorption. Desorption is not, however, the only consequence from the interaction between cosmic rays and dust grains. In fact, recent experiments with interstellar ices (Dartois et al. 2015) show that cosmic irradiation can also alter the ice mantle state on dust grains.

The cosmic-ray reaction rates on grain surfaces, $R$$_{\mathrm{CR}}$ (cm$^{-3}$s$^{-1}$), are assumed to be the same as the rates for gas-phase reactions. They are determined by

\begin{equation}
\kappa_{\mathrm{CR}}= z_{{i}} \zeta_{H_{2}},
\end{equation}

\begin{equation}
R_{\mathrm{CR}}= n_{{i}} \kappa_{\mathrm{CR}},
\end{equation}

\noindent where $\kappa$$_{\mathrm{CR}}$ (s$^{-1}$) is the cosmic-ray rate coefficient, $n$$_{\mathrm{i}}$ is the number density of the photodissociated species, $z$$_{\mathrm{i}}$ is the cosmic-ray ionisation rate factor, which depends on the ionising element (see KIDA database and Table \ref{table:cosmic_ray_reactions}), and $\zeta$$_{\mathrm{H_2}}$ is the cosmic-ray ionisation rate per H$_2$ molecule ($\zeta$$_{\mathrm{H_2}}$=5$\times$10$^{-17}$ s$^{-1}$; Indriolo et al. 2007; Hocuk $\&$ Spaans 2011). The cosmic-ray reactions on dust grains considered in this code are listed in Table \ref{table:cosmic_ray_reactions}.

\subsection{Dust temperature}
\label{Dust_temperature}

The dust temperature is a key parameter in the thermal balance calculation, since it influences the gas temperature, through heating and cooling rates along with chemical reaction rates. In addition, the dust temperature value is also crucial for the formation of ice mantles on grain surfaces.
There are several expressions for dust temperature in the literature, such as those derived by Werner $\&$ Salpeter (1969), Hollenbach et al. (1991), Zucconi et al. (2001), and Garrod $\&$ Pauli (2011). See also Hocuk et al. (in prep.), who show a fit of dust temperature observations as a function of visual extinction, using the different analytical expressions of $T$$_{\mathrm{d}}$ previously mentioned. 

In our code, we considered the most recent dust temperature expression derived by Garrod $\&$ Pauli (2011), but with an adaptation (private communication with Garrod), that includes a dependence on radiation field, since the original expression only depends on the visual extinction ($A$$_{\mathrm{V}}$). The final expression is

\begin{eqnarray}
T_{\mathrm{d}} = 18.67 - 1.637  \left(\begin{array}{c}A_{\mathrm{V}} - \mathrm{log}(G_{0}) \end{array} \right) + \nonumber \\
  0.07518 \left(\begin{array}{c}A_{\mathrm{V}} - \mathrm{log}(G_{0}) \end{array} \right)^2
- 0.001492 \left( \begin{array}{c}A_{\mathrm{V}} - \mathrm{log}(G_{0}) \end{array} \right)^3.
\end{eqnarray}\\


Dust grains in strong radiation field environments present not only high temperatures, but also grain temperature fluctuations, as derived by Cuppen et al. (2006) and Iqbal et al. (2014) using Monte Carlo simulations, and by Bron et al. (2014) using an analytical approach. In this paper, however, we have not included a formalism taking temperature fluctuations of a grain size distribution into account, since we have large grains ($\gtrsim$50 \AA $\ $because of the considered MRN distribution) and large fluctuations mainly occur for smaller grains (Draine  $\&$  Li 2001){\footnote{https://ned.ipac.caltech.edu/level5/March04/Draine/Figures/figure5.jpg}}.

\section{Results}
\label{Results}

\begin{figure}[h!]
\begin{center}
   \includegraphics[angle=0,width=6.7cm]{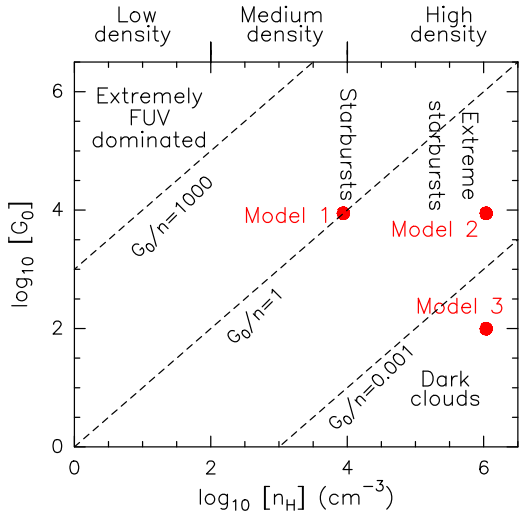} 
   \caption{Diagram indicating different regimes in the ($n$$_{\mathrm{H}}$, $G$$_{0}$) parameter space (adapted from Kazandjian et al. 2015). The red points correspond to our models.}
   \label{figure:Environments}
   \end{center}
   \end{figure}

In this section, we discuss the results for three models in which we varied the radiation field\footnote{We use $G$$_0$, the Habing field (Habing 1968), as the normalisation in which we express the incident FUV radiation field, where $G$$_0$=1 corresponds to a flux of 1.6$\times$10$^{-3}$ erg cm$^{-2}$ s$^{-1}$.} ($G$$_0$) and the density  ($n$$_{\mathrm{H}}$) for a semi-infinite slab geometry and irradiation from one side without geometrical dilution. The adopted model parameters are listed in Table \ref{table:parameters} (see also Fig. \ref{figure:Environments}, which indicates the regions of the $G$$_0$-$n$$_{\mathrm{H}}$ space occupied by the three models and various astrophysical objects). In Model 1, we study a typical photodissociation region (e.g. the Orion Bar) characterised by high density and strong radiation field conditions. In Model 2 (typical conditions of an extreme starburst) and Model 3, we study the consequences of varying the density and intensity of the radiation field, respectively.   

\begin{table}[h!]
\caption{Adopted model parameters.}             
\centering   
\begin{tabular}{l l l l}     
\hline\hline       
Model & $G$$_{0}$   & $F$$_{\mathrm{FUV}}$     & $n$$_{\mathrm{H}}$ \\ 
      &             & (erg cm$^{-2}$ s$^{-1}$) &  (cm$^{-3}$) \\
\hline                    
1  & 10$^4$      & 16    & 10$^{4}$   \\
2  & 10$^{4}$    & 16    & 10$^{6}$   \\
3  & 10$^2$      & 0.16  & 10$^{6}$    \\
\hline
\label{table:parameters}                  
\end{tabular}
\end{table}

\subsection{Heating and cooling}

\begin{figure}
\includegraphics[scale=0.4, angle=0]{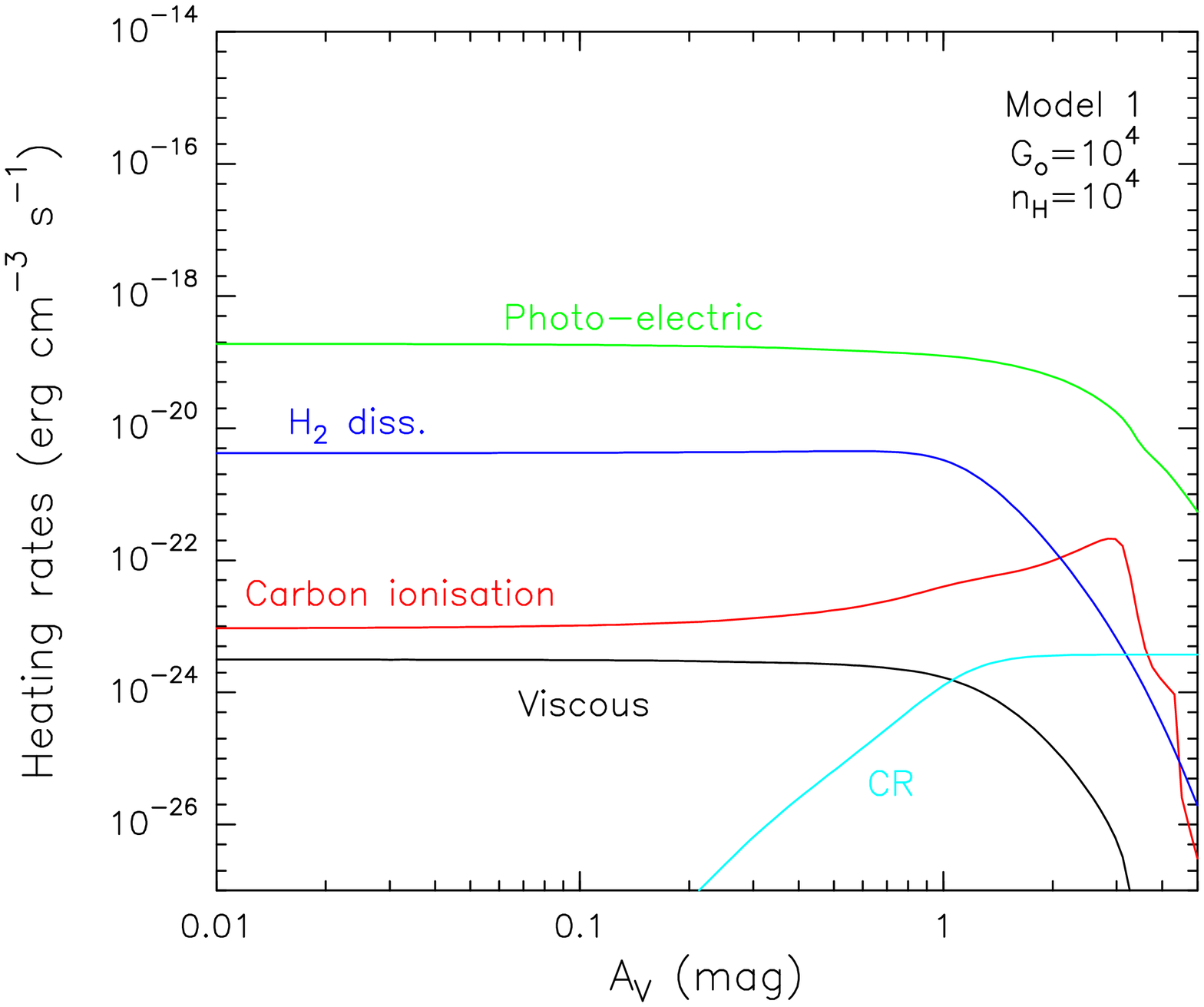} \vspace{0.5cm}\\
\vspace{0.5cm}
\includegraphics[scale=0.4, angle=0]{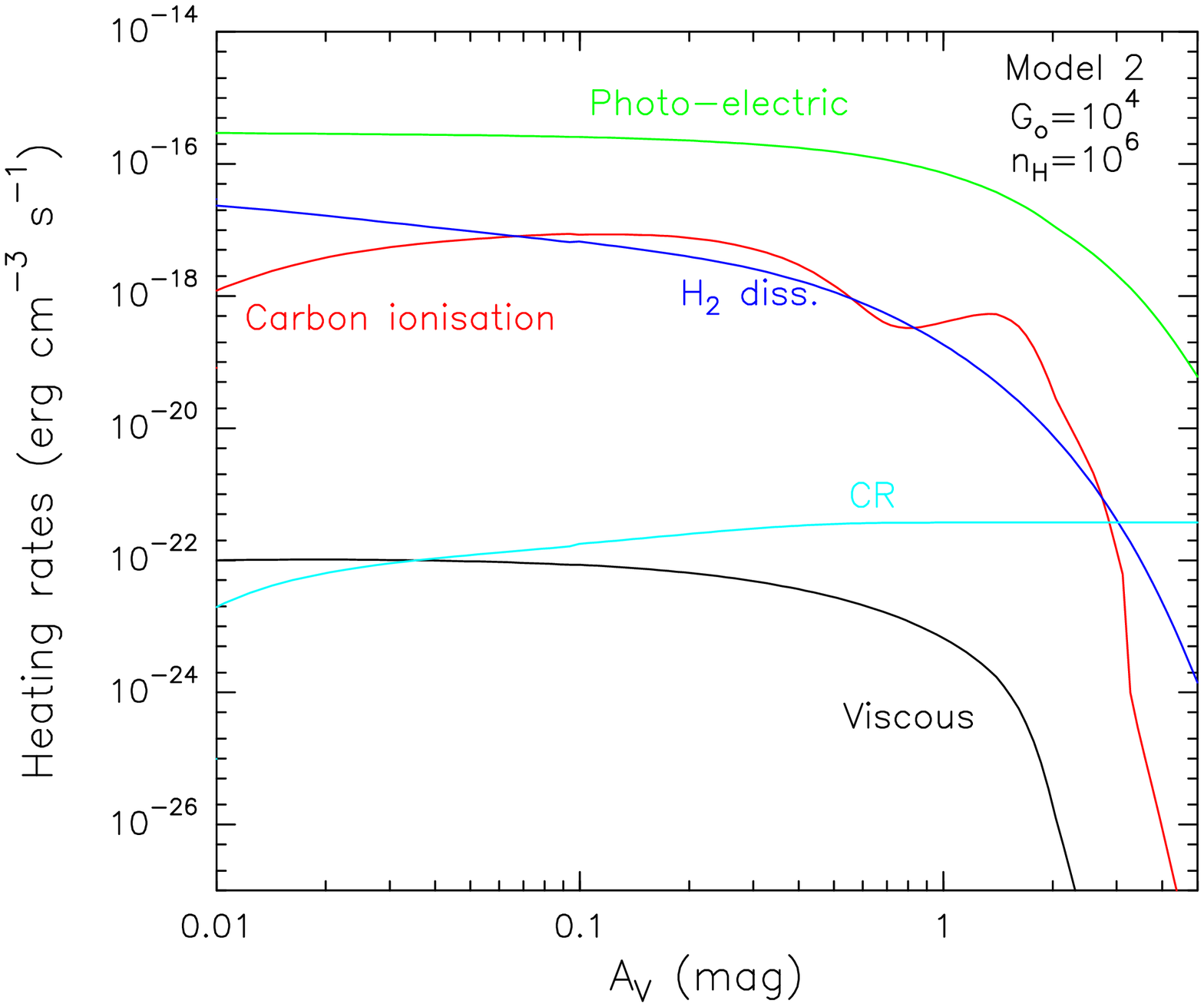}\\
\vspace{0.5cm}
\includegraphics[scale=0.4, angle=0]{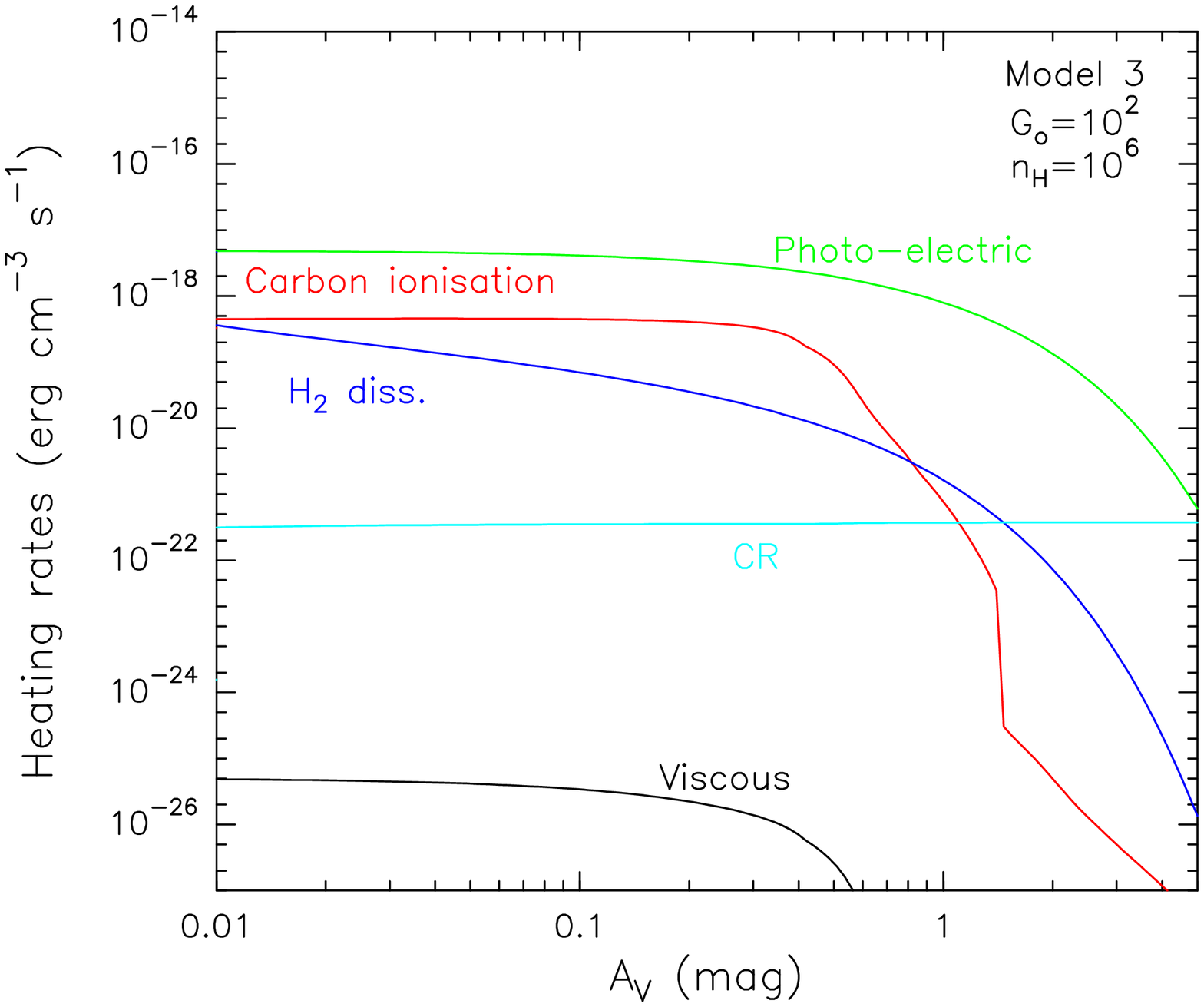} \\
\caption{Most important heating processes for Models 1, 2, and 3.}
\label{figure:heating_rates}
\end{figure}

\begin{figure}
\includegraphics[scale=0.415, angle=0]{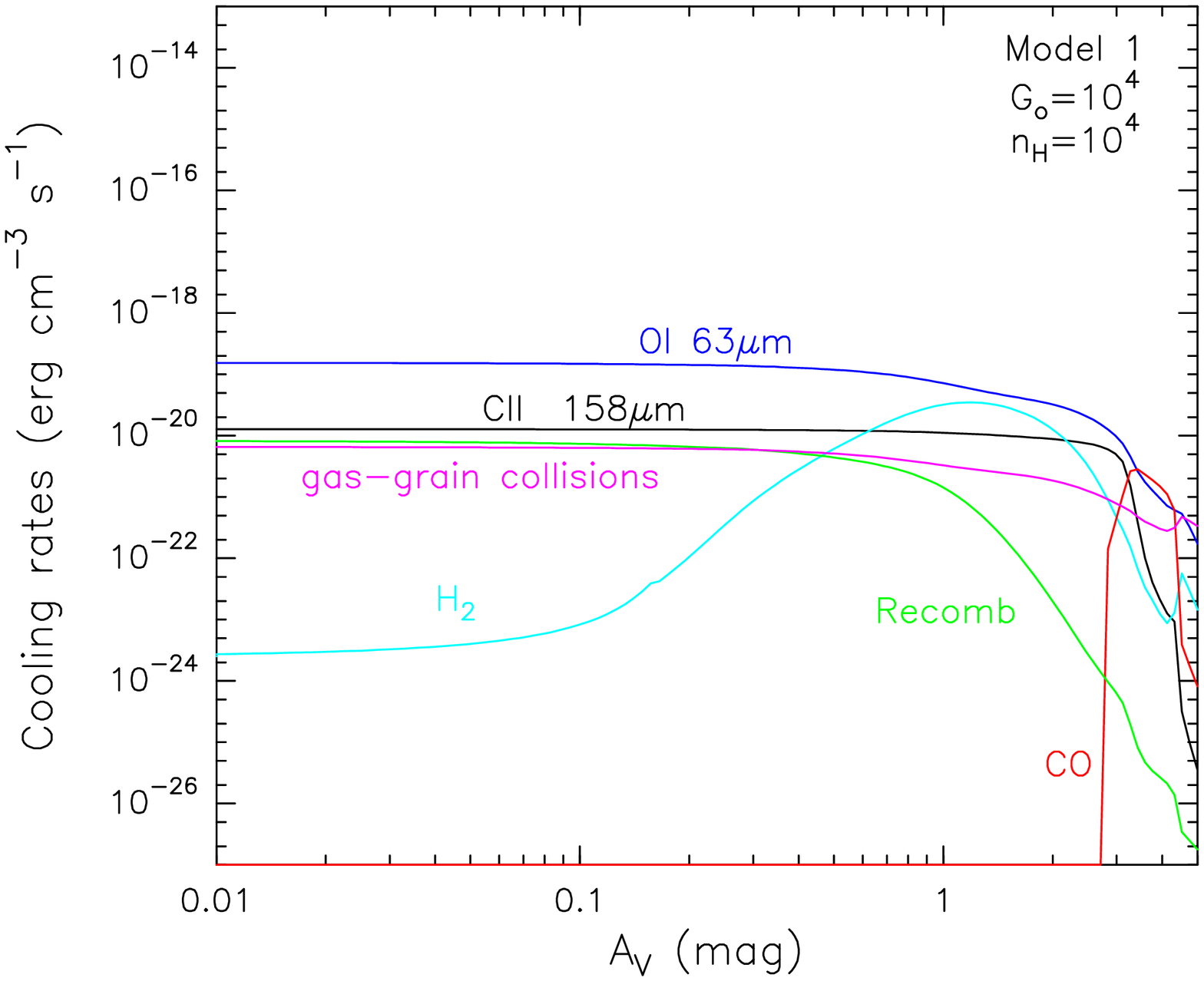} \vspace{0.5cm}\\
\vspace{0.5cm}
\includegraphics[scale=0.415, angle=0]{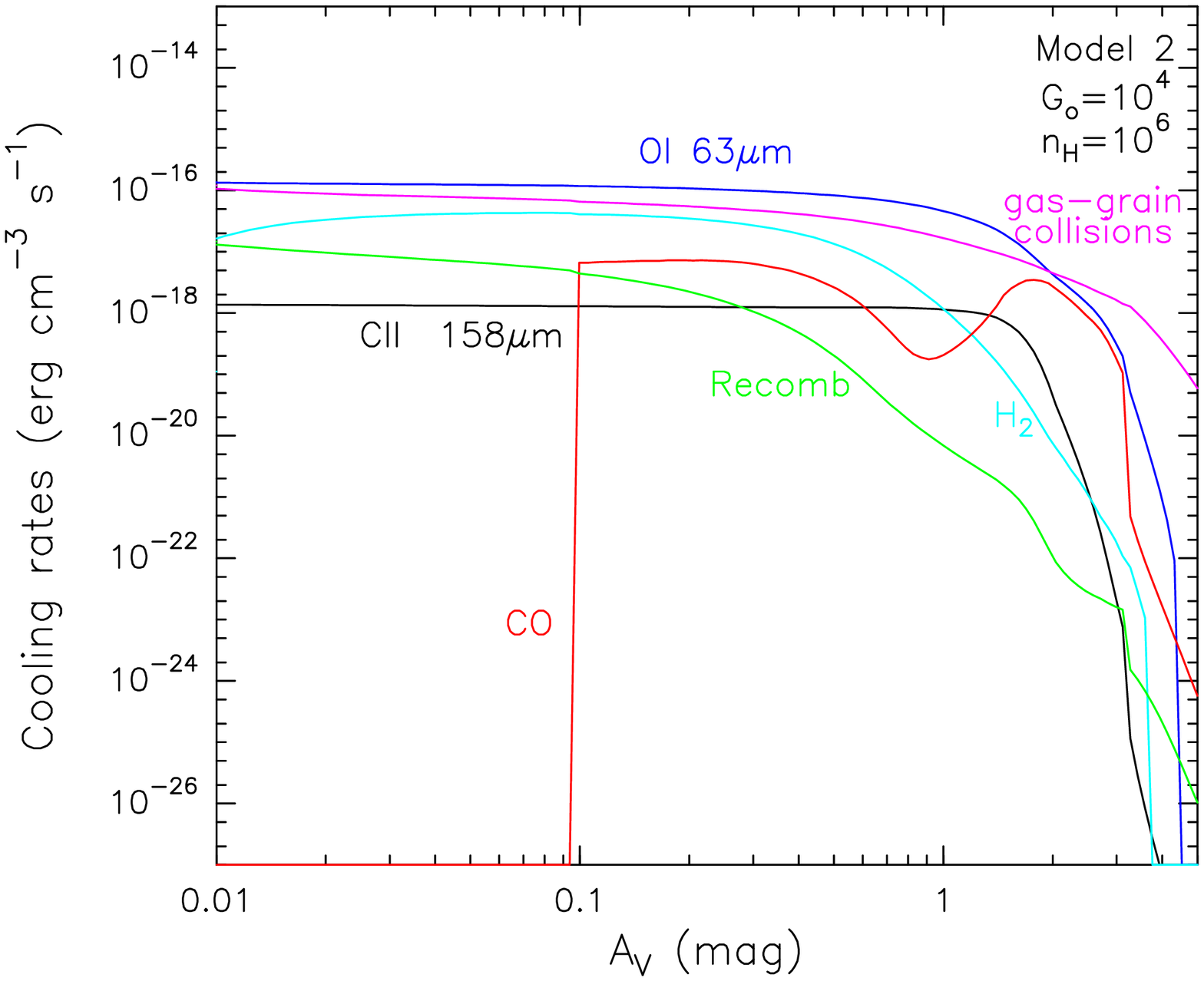}\\
\vspace{0.5cm}
\includegraphics[scale=0.415, angle=0]{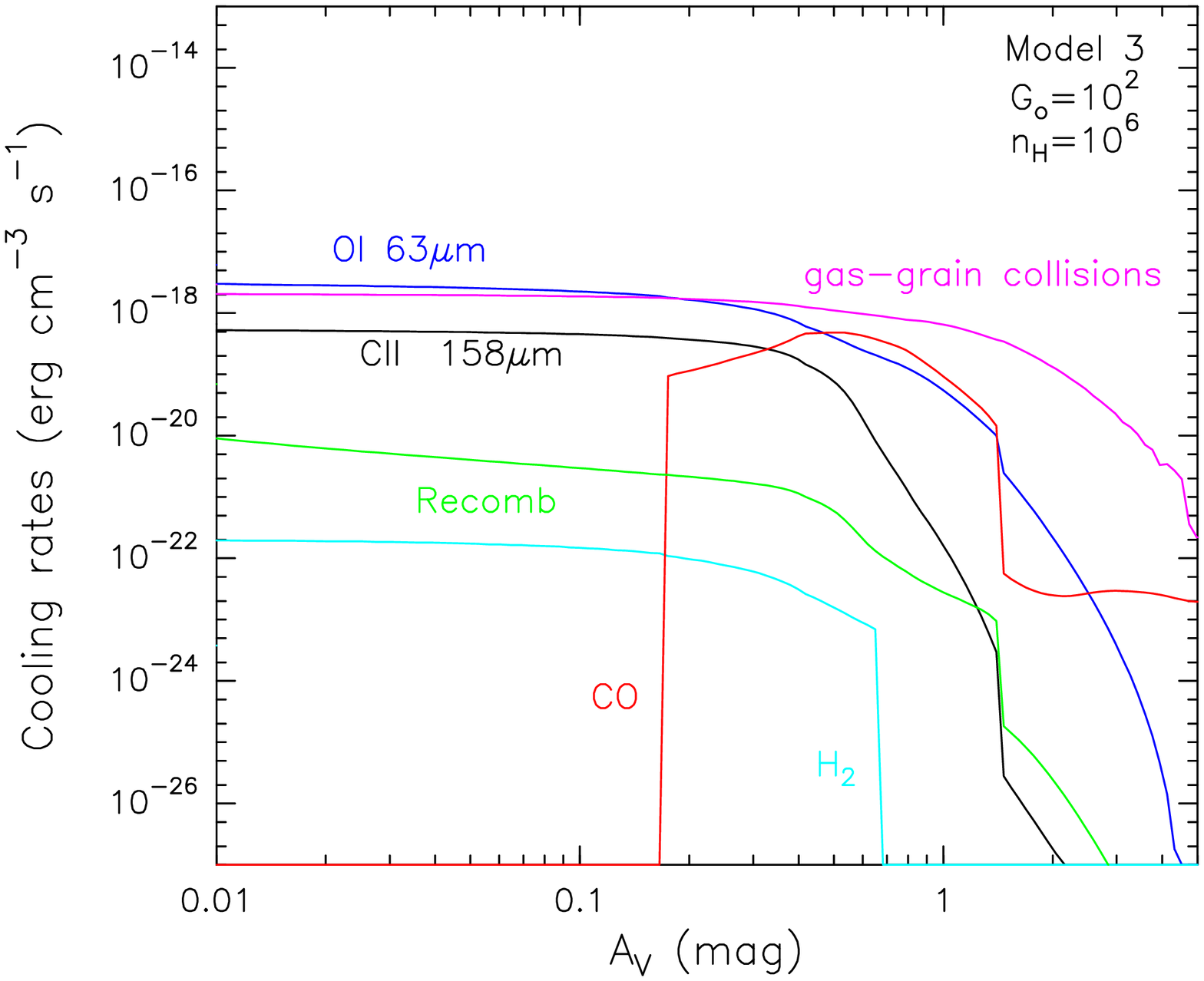} \\
\caption{Most important cooling processes for Models 1, 2, and 3.}
\label{figure:cooling_rates}
\end{figure}

In Sect. \ref{Gas_chemistry}, we listed the different heating and cooling mechanisms considered in our code. The results for the most relevant heating rates for Models 1-3 are shown in Fig. \ref{figure:heating_rates} as a function of the visual extinction. For both radiation fields and densities considered (see Table \ref{table:parameters}), the dominant heating source for visual extinctions $A$$_{\mathrm{V}}$$<$5 mag is photoelectric emission from grains. At $A$$_{\mathrm{V}}$$<$0.5 mag, if the intensity of the radiation field is high (Model 1 and 2), we also obtain a high contribution from the H$_2$ photodissociation heating. By contrast, if $G$$_0$ is low (Model 3) the second highest contribution to the heating arises from carbon ionisation. Deeper in the cloud, $A$$_{\mathrm{V}}$$>$2 mag, the heating by photoelectric emission, although still dominant, becomes progressively less efficient. For a model with high density and low $G$$_0$ (Model 3), cosmic rays dominate the heating of the region together with photoelectric emission processes at $A$$_{\mathrm{V}}$$\sim$5 mag. 
We also find that viscous heating never contributes significantly to the heating and that gas-grain collisions act as a cooling mechanism.

In Fig. \ref{figure:cooling_rates}, we show the cooling rates for Models 1-3 as a function of the visual extinction. For a PDR with intermediate density ($n$$_{\mathrm{H}}$=10$^4$ cm$^{-3}$, Model 1), the cooling is dominated by [OI] 63 $\mu$m at $A$$_{\mathrm{V}}$$\lesssim$5 mag. For a higher density PDR (Models 2 and 3), the cooling is dominated by [OI] 63 $\mu$m and gas-grain collisions up to $A$$_{\mathrm{V}}$$\sim$2 mag and $\sim$0.5 mag, respectively. For higher extinctions, however, [OI] 63 $\mu$m becomes inefficient and gas-grain collisions represent the main coolant. Other processes, such as recombination of electrons with grains, represent minor coolants, especially at $A$$_{\mathrm{V}}$$>$1 mag.

 \subsection{Chemical structure}
\label{chemical_structure}

As stated previously, the physical and chemical processes in PDRs are dominated by interactions with photons, whose timescales are much lower than those for dynamical processes occurring in the opaque interiors of the clouds (Hollenbach et al. 2009). At very large visual extinctions, steady-state chemical codes, such as our PDR code, do not apply because of the need of considering the time-dependent chemical network, since certain chemical timescales are comparable to cloud lifetimes ($\sim$10$^7$ years). Therefore, to analyse the chemical structure of different types of PDRs, we consider results for low visual extinctions ($A$$_{\mathrm{V}}$$\lesssim$1 mag), where physical and chemical processes are purely dominated by interactions with photons, and for extinctions of translucent clouds ($A$$_{\mathrm{V}}$$\sim$1-5 mag).

\subsubsection{Gas-phase species}
\label{gas-phase}

\begin{figure}
\hspace{-0.5cm}
\includegraphics[scale=0.32, angle=0]{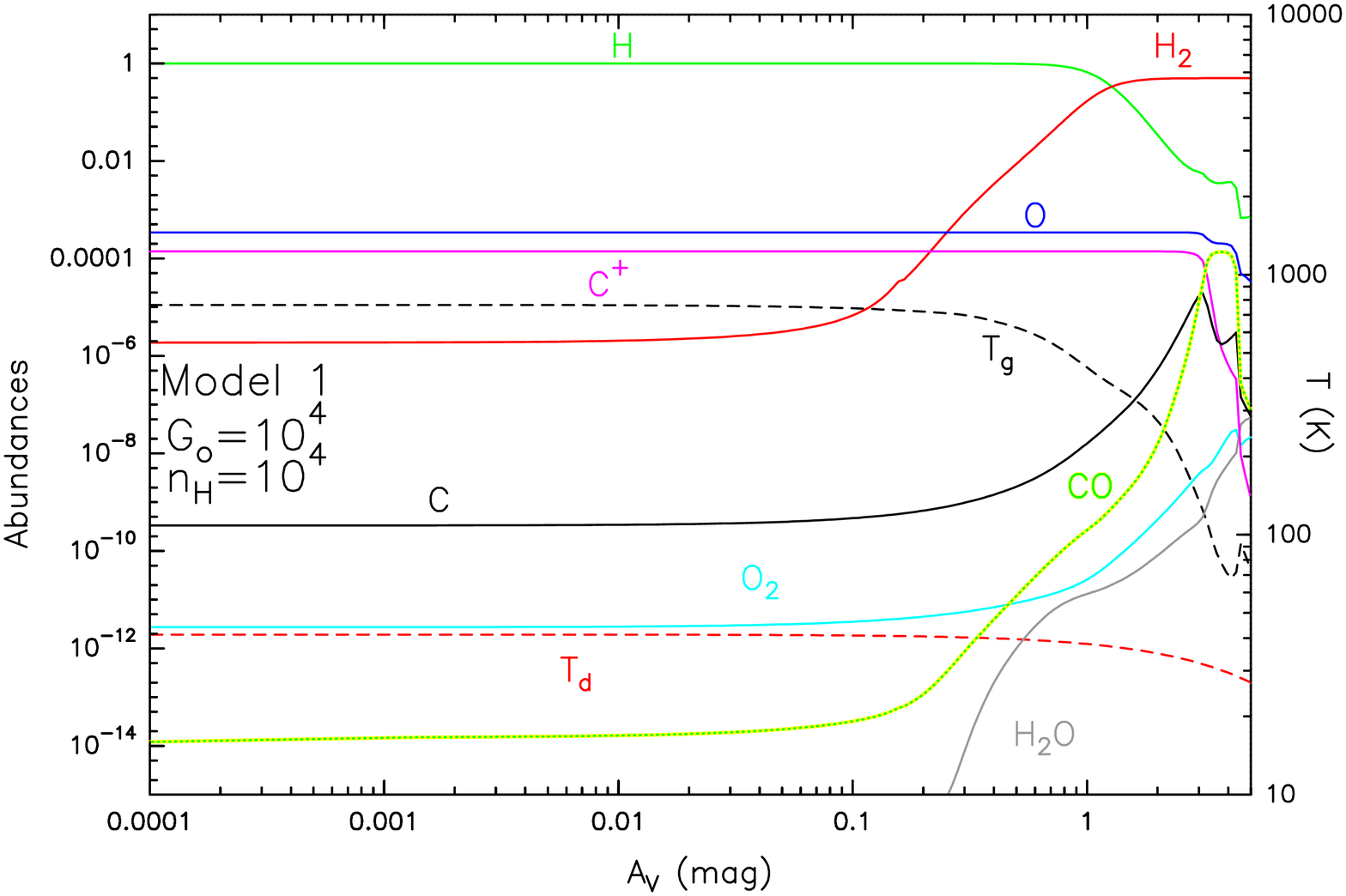} \vspace{0.5cm} \\
\vspace{0.5cm}
\hspace{-0.5cm}
\includegraphics[scale=0.32, angle=0]{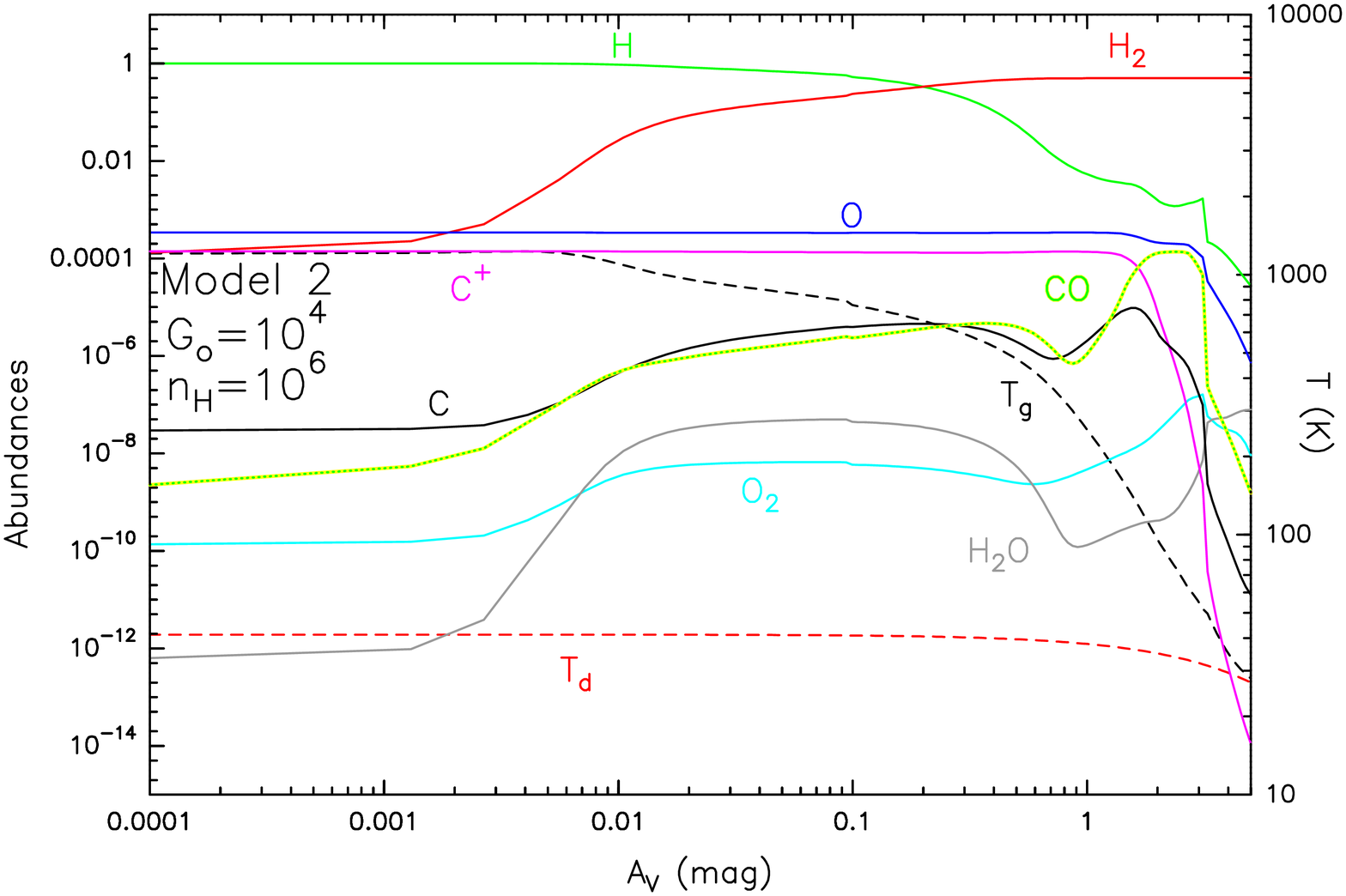}\\
\vspace{0.5cm}
\hspace{-0.5cm}
\includegraphics[scale=0.32, angle=0]{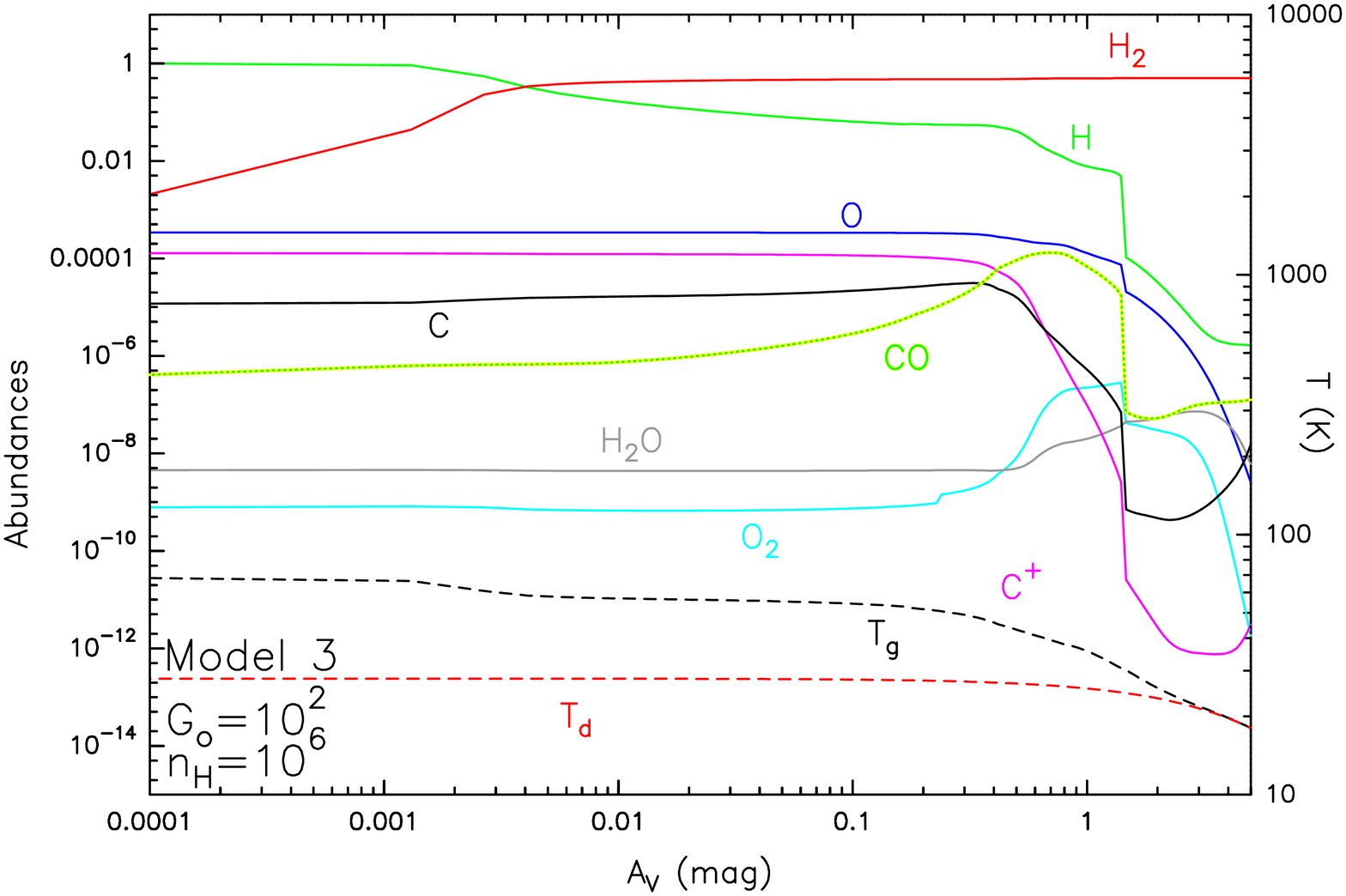} \\
\hspace{-0.5cm}
\caption{Fractional abundances, $n$(x)/$n$$_{\mathrm{H}}$, of gas species for Models 1, 2, and 3.}
\label{figure:abundances_results_gas}
\end{figure}

In Fig. \ref{figure:abundances_results_gas}, we show the fractional abundances of several gas-phase species as a function of the visual extinction for Models 1, 2, and 3. We also plot the dust temperature ($T$$_{\mathrm{d}}$) and the gas temperature ($T$$_{\mathrm{g}}$) for each model. 
We obtain a H$\rightarrow$H$_2$ transition with its location varying significantly depending on the radiation field and density. The atomic hydrogen is converted into H$_2$ at deeper locations in the cloud for high radiation fields (Models 1 and 2), since the photodissociation rates are larger. With a low $G$$_0$ and high density model (Model 3), however, the transition occurs closer to the cloud surface, since the chemical rates depend on $n$$^2$. We also observe that the H-H$_2$ transition becomes sharper as the intensity of the radiation field increases and the density decreases, in agreement with analytical results obtained by Sternberg et al. (2014).

\begin{figure}[h!]
\vspace{0.3cm}
\hspace{0.9cm}
\includegraphics[scale=0.385, angle=0]{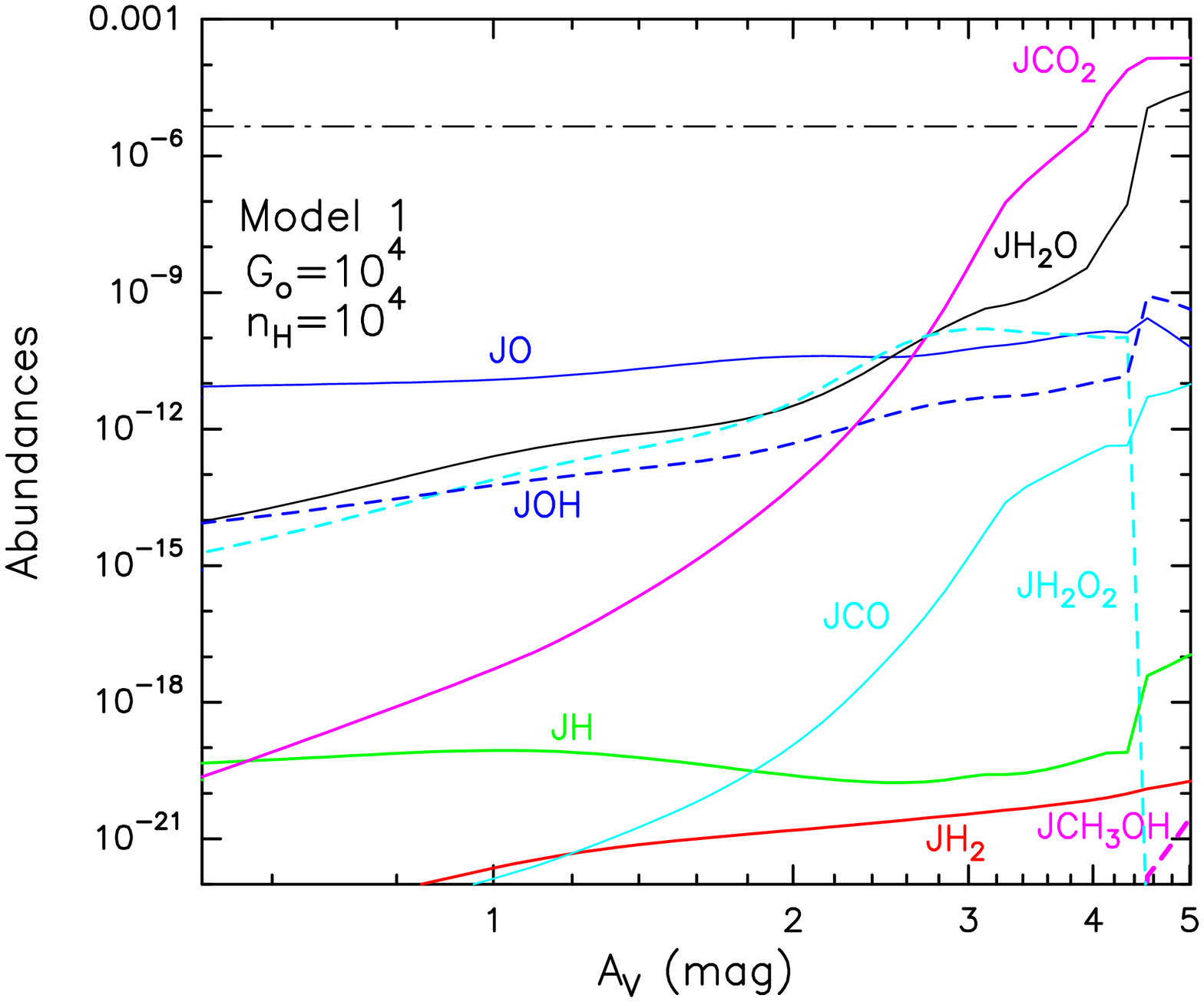} \vspace{0.5cm} \\
\vspace{0.7cm}
\hspace{0.9cm}
\includegraphics[scale=0.385, angle=0]{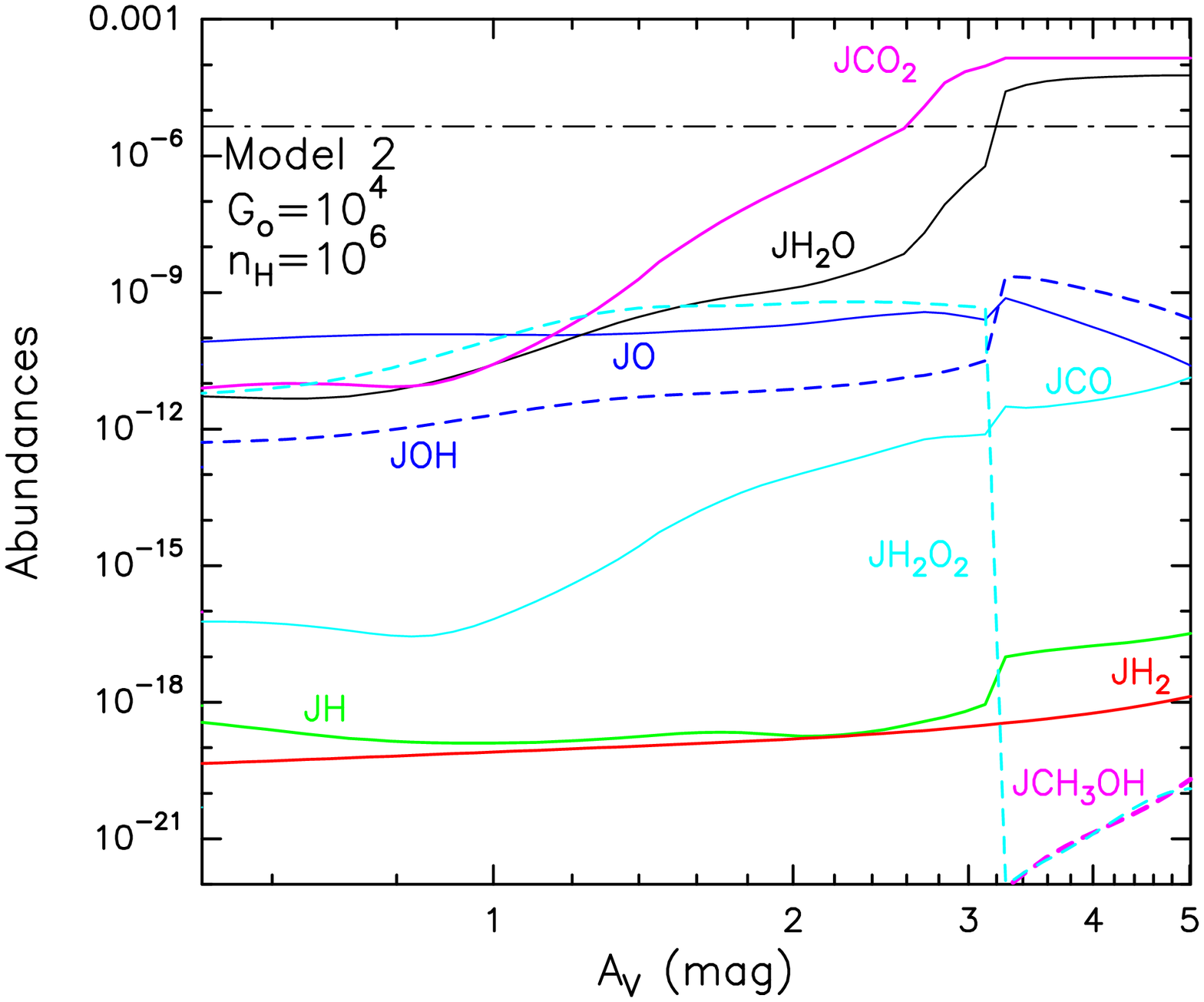}\\
\vspace{0.5cm}
\hspace{0.9cm}
\includegraphics[scale=0.385, angle=0]{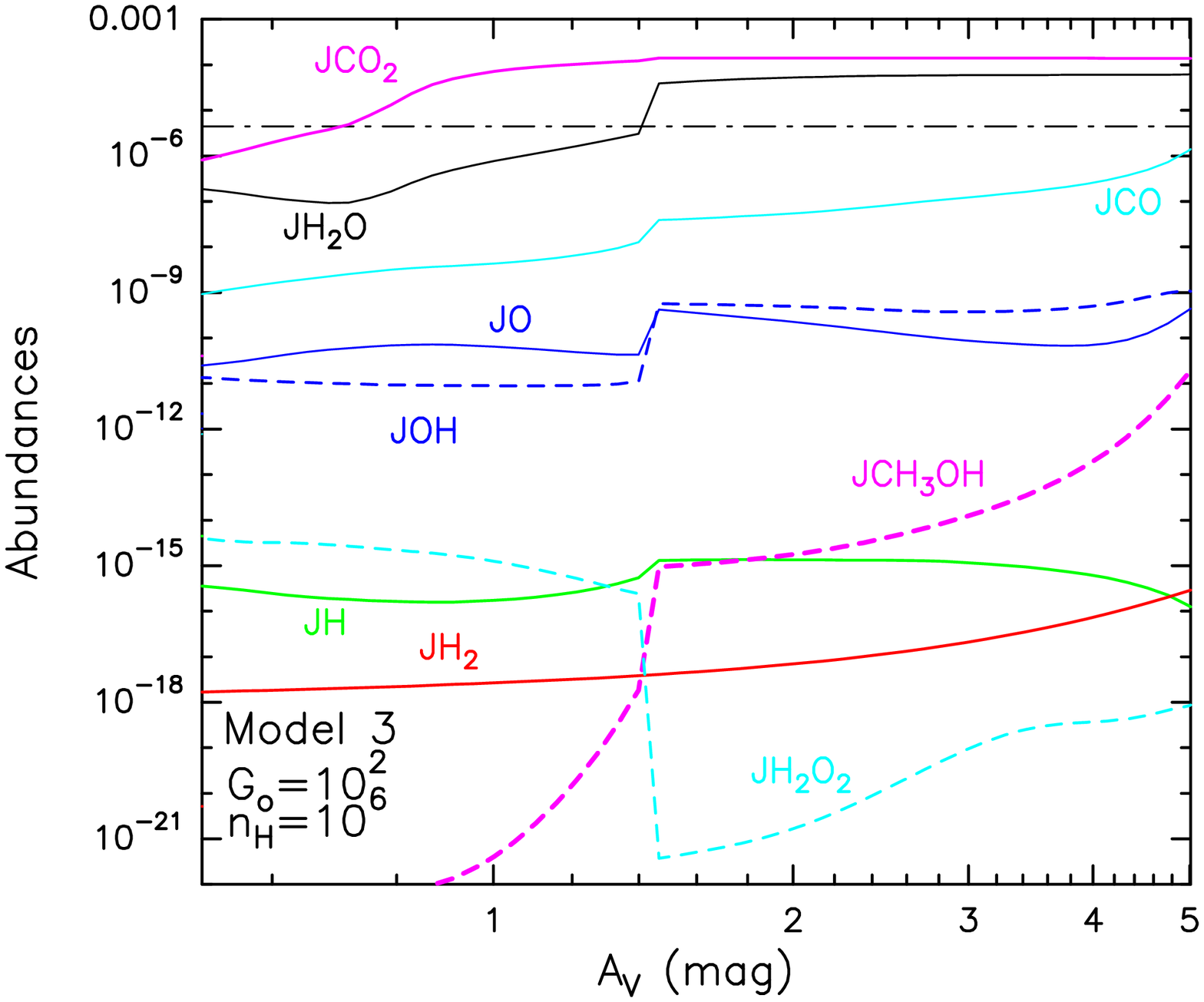} \\
\caption{Fractional abundances, $n$(x)/$n$$_{\mathrm{H}}$, of ice species for Models 1, 2, and 3. JX means solid X. The dash-dotted black line represents the number of possible attachable sites on grain surfaces per cm$^{3}$.}
\label{figure:abundances_results_dust}
\end{figure}

Unlike results obtained by Meijerink \& Spaans (2005) and with other PDR codes (which only consider gas-phase chemistry), where C$^+$ presents high abundances at low visual extinctions, C at intermediate extinctions, and CO at high extinctions, here the transition C$^+$$\rightarrow$C$\rightarrow$CO is no longer well-defined. 
While the abundance of C$^+$ is high at low extinctions for all models and it decreases as the cloud becomes denser, at intermediate extinctions ($A$$_{\mathrm{V}}$$\sim$2 mag) for Models 1 and 2, and at $A$$_{\mathrm{V}}$$\sim$0.5 mag for Model 3, most of the atomic carbon is rapidly converted into CO leading to low C abundances.
Similar results for atomic carbon are found by Hollenbach et al. (2009) in their steady-state PDR code modelling the formation of CO and H$_2$O ices, which suggests that surface chemistry is accelerating the formation of CO. 
In Models 2 and 3, we also observe how the O abundance dramatically decreases by several orders of magnitude at $A$$_{\mathrm{V}}$$\sim$3 mag. 
These decreases are due to the oxygen depletion, which is locked in water ice. This fast increase of the solid water abundance can be seen in Fig. \ref{figure:abundances_results_dust} (Sect. \ref{dust-phase}).

\subsubsection{Dust-phase species}
\label{dust-phase}

\begin{figure}
\vspace{0.3cm}
\hspace{0.9cm}
\includegraphics[scale=0.39, angle=0]{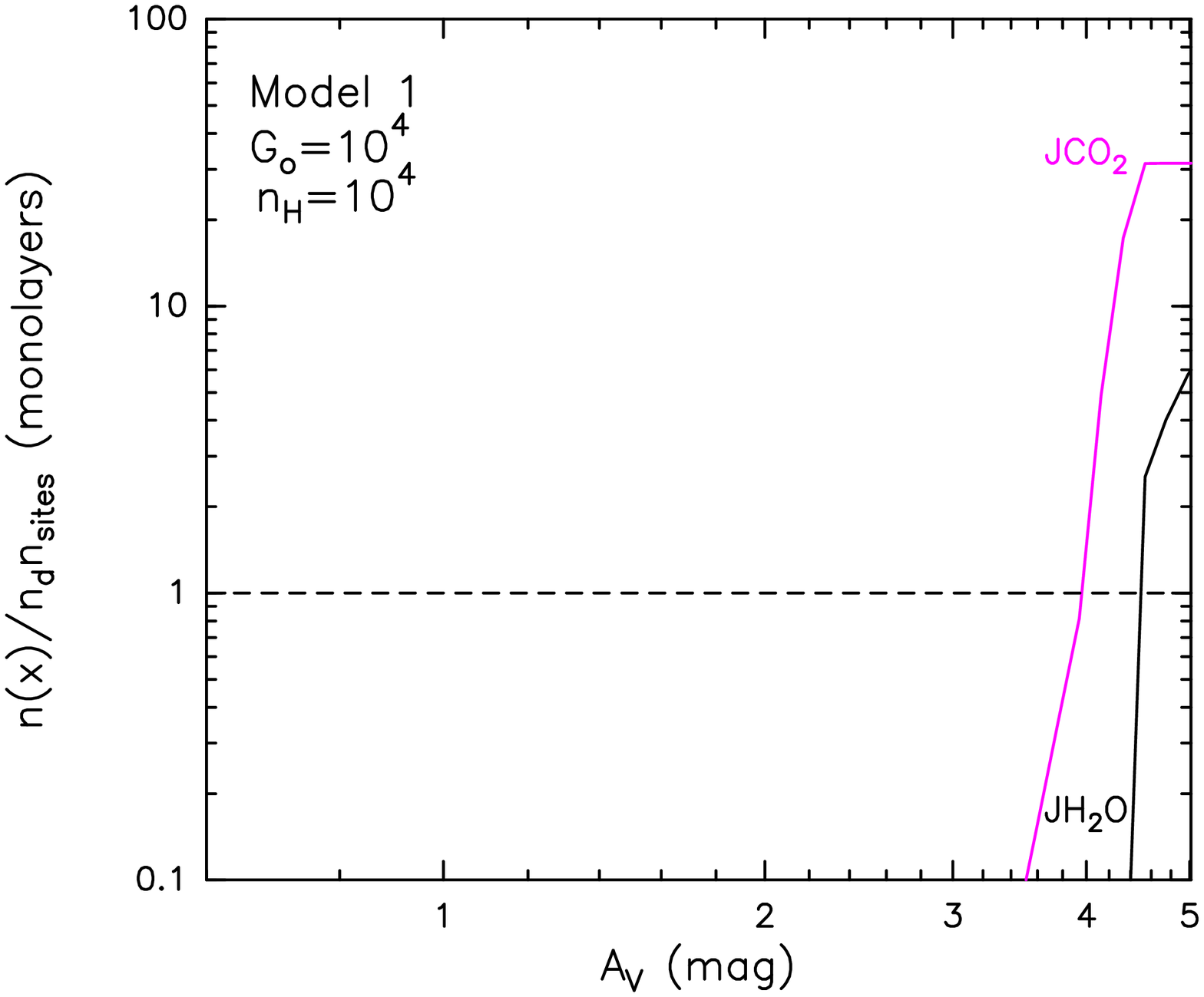} \vspace{0.5cm} \\
\vspace{0.7cm}
\hspace{0.9cm}
\includegraphics[scale=0.39, angle=0]{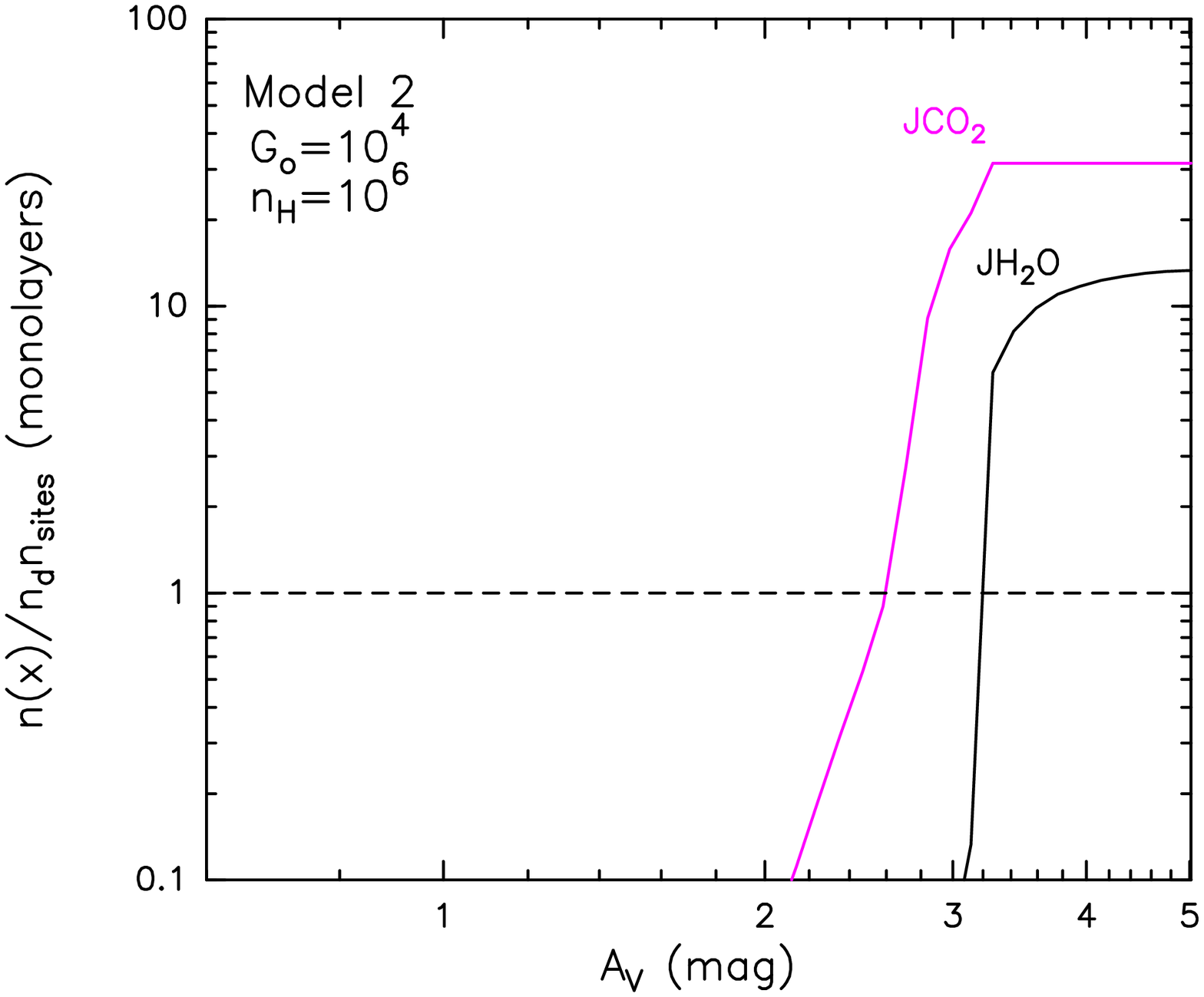}\\
\vspace{0.5cm}
\hspace{0.9cm}
\includegraphics[scale=0.39, angle=0]{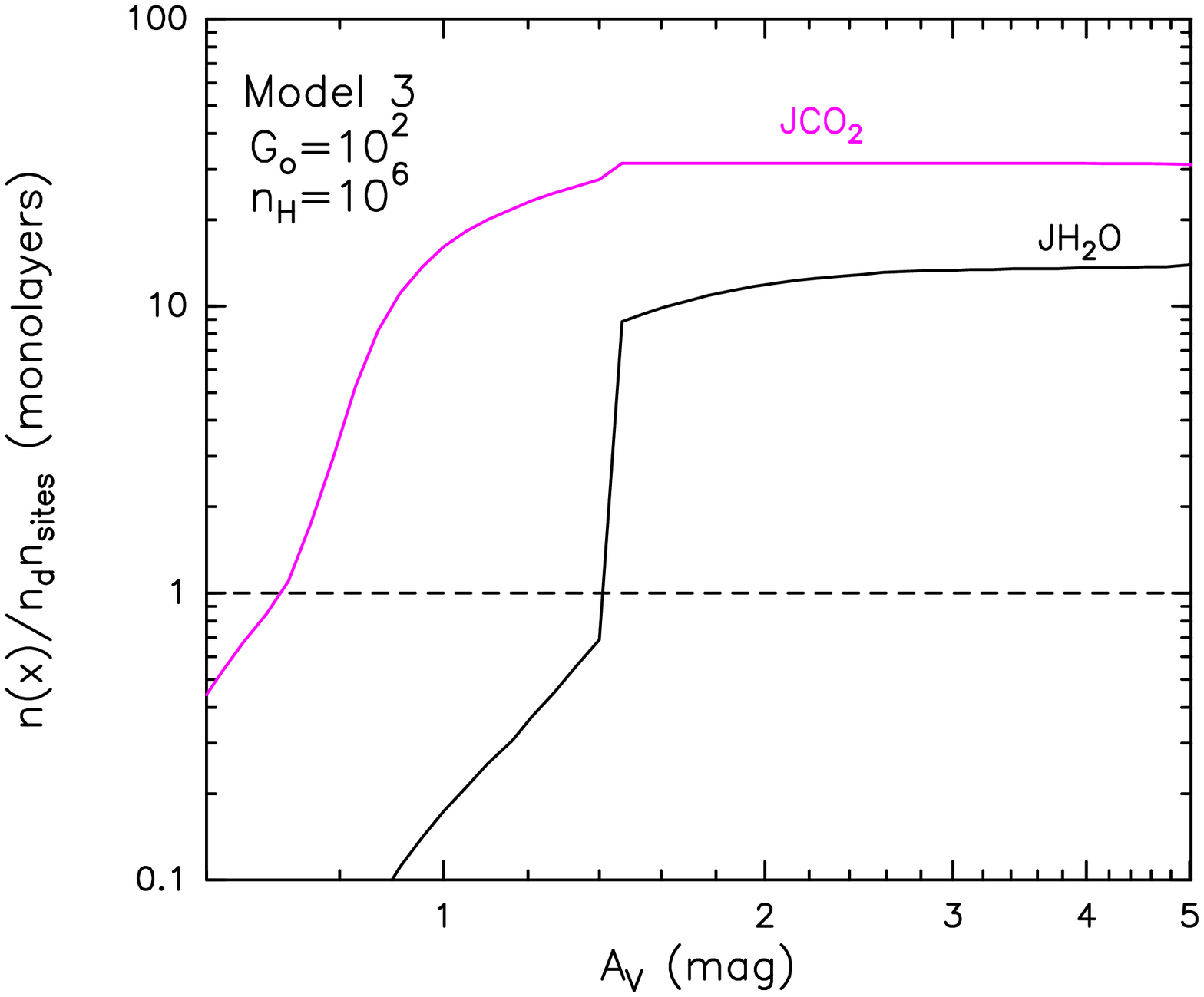} \\
\caption{Growth of ice layers on grain surfaces for H$_2$O and CO$_2$ in Models 1 (top), 2 (middle), and 3 (bottom). JX means solid X.}
\label{figure:monolayers}
\end{figure}

  In Fig. \ref{figure:abundances_results_dust}, we show the fractional abundances of several solid species as a function of the visual extinction for Models 1, 2, and 3. The dash-dotted black line represents the number of possible attachable sites on grain surfaces per cm$^{3}$ of space (see expression 4), i.e. the limit to reach a full monolayer of ice. We observe that the solid CO$_2$ and H$_2$O abundances in all of the models become larger than this limit, leading to the formation of several ice layers of these two species (see Fig. \ref{figure:monolayers}, which describes the exact number of ice layers of CO$_2$ and H$_2$O that are formed).

The radiation field strength is one of the most important factors governing ice formation, since it determines the dust temperature. Stronger radiation fields lead to higher dust temperature and prevent ice formation until large depths. We observe this effect comparing results from Models 2 and 3. In Model 3, with a low $G$$_0$ (10$^{2}$), the dust temperature at $A$$_{\mathrm{V}}$$\sim$5 mag is $<$15 K, while in Model 2, $T$$_{\mathrm{d}}$$\sim$25 K owing to the higher intensity of the radiation field in this case ($G$$_{0}$=10$^4$). This leads to the CO$_2$ ice formation at lower $A$$_{\mathrm{V}}$ ($\sim$0.7 mag) in Model 3 compared to Model 2 ($A$$_{\mathrm{V}}$$\sim$2.5 mag). Similar results are found for H$_2$O ices; the formation of a full water ice monolayer takes places at $\sim$2 mag lower in Model 3 (low $G$$_{0}$) than in Model 2 (high $G$$_{0}$).

The density also plays an important role in the ice formation processes, since it affects the number of ice layers that are formed for water (see Fig. \ref{figure:monolayers}, Models 1 and 2). In particular, in the PDR with the lowest density (Model 1), $\sim$6 ice monolayers of H$_2$O are formed at $A$$_{\mathrm{V}}$$\leq$5 mag against the $\sim$15 ice monolayers formed when the density is increased by two orders of magnitude. Comparing Models 2 and 3 in Fig. \ref{figure:monolayers}, we also conclude that the variation of the radiation field intensity barely affects the number of formed monolayers of CO$_2$ and H$_2$O ice. 
In a low radiation and high density PDR (Model 3, bottom pannel of Fig. \ref{figure:monolayers}), we find the formation of the first monolayer of CO$_2$ and H$_2$O ice at very low extinctions ($A$$_{\mathrm{V}}$$\lesssim$1.5 mag). This is due to the shielding effect produced by the high density of the region, which prevents ice destruction by the impact of UV photons. In particular, in Model 3 ($G$$_0$=100 and $n$=10$^6$ cm$^{-3}$) we find the H$_2$O ice threshold extinction at $A$$_{\mathrm{V}}$$\sim$1.5 mag. Recent observational results of the IC 5146 dark coud ($n$$\sim$10$^5$ cm$^{-3}$) and of low-mass young stellar objects show the H$_2$O ice threshold extinction at $A$$_{\mathrm{V}}$$\sim$3 mag, which is equivalent to that found for the Taurus dark cloud (Chiar et al. 2011, Noble et al. 2013). Taking into account our results showing that a decrease of density in two orders of magnitude leads to an increase of the threshold extinction to form ices of $\sim$1.5 mag, our model results are in agreement with observations. For CO, the ice threshold is significantly higher ($A$$_{\mathrm{V}}$$\sim$5-11 mag), according to observations of Taurus and $\rho$ Ophiuchi (Whittet et al. 1989; Shuping et al. 2000; Velusamy et al. 2005).

In the previous Section, it was mentioned that the abundance of solid water increases as the gas-phase oxygen is depleted. In Fig. \ref{figure:abundances_results_dust} (Models 2 and 3), we also observe that at a given high visual extinction (different for each model), the water ice growth slows down, while the gas-phase oxygen abundance keeps decreasing. This is because the ice abundance saturation after most of the O nuclei are locked in water ice. 
From results in Fig. \ref{figure:abundances_results_dust}, we also deduce that a low radiation field promotes the formation of solid methanol, since the surface reaction between H and H$_3$CO becomes more efficient as the visual extinction increases. 
Regarding to solid H$_2$O$_2$, we obtain the highest fractional abundances to be $\sim$$10$$^{-14}$ for a low $G$$_0$ PDR and $\sim$$10$$^{-10}$ for a high $G$$_0$ PDR in agreement with results from Ioppolo et al. (2008). After reaching its maximum abundances, we observe that solid H$_2$O$_2$ decreases sharply, mainly owing to its destruction by reacting with solid H to form solid water.

   
\section{Discussion}
\label{Discussion}

\subsection{Ice species formation rates}
\label{Rates}

\subsubsection{H$_2$O ice}
\label{H2O_ice}

\begin{figure}
   \centering
\includegraphics[angle=0,width=9.1cm]{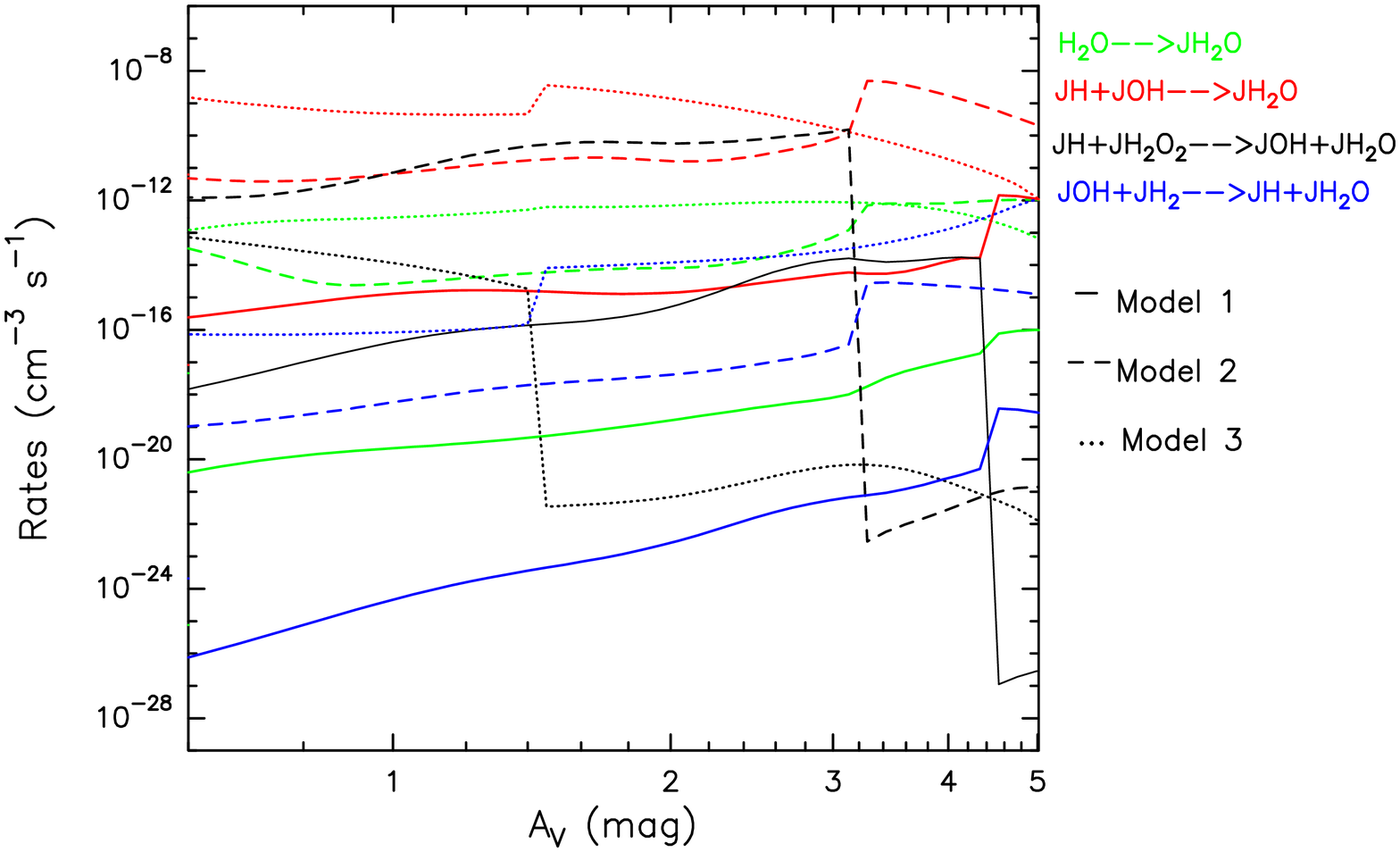}
   \caption{Rates for surface reactions forming H$_2$O ice. JX means solid X.}          
   \label{figure:H2O_ice_formation_rates}
   \end{figure}

In Fig. \ref{figure:H2O_ice_formation_rates}, we show rates for the main surface reactions forming solid water. We obtain that H$_2$O$_2$ on dust grains is an important intermediate in the formation of solid water for a high $G$$_0$ PDR (Models 1 and 2) at $A$$_{\mathrm{V}}$$\lesssim$3 mag, in agreement with Du et al. (2012). However, the main surface reaction leading to water ice monolayers at larger extinctions is the reaction between solid H and OH. By contrast, for a low $G$$_0$ PDR (Model 3), the rates of this reaction decreases as the visual extinction increases, until the reaction between solid OH and H$_2$ becomes equally important in the water ice formation.
Other surface reactions present rates that are too low (mainly due to the low abundances of some of the solid reactant) to significantly contribute to the formation of water ice at $A$$_{\mathrm{V}}$$\lesssim$5 mag. This is, for example, the case of the surface reaction OH+CH$_3$OH$\rightarrow$H$_3$CO+H$_2$O. We also find a small contribution to solid water formation from water depletion in all the models.

  \begin{figure}
   \centering
\includegraphics[angle=0,width=9.1cm]{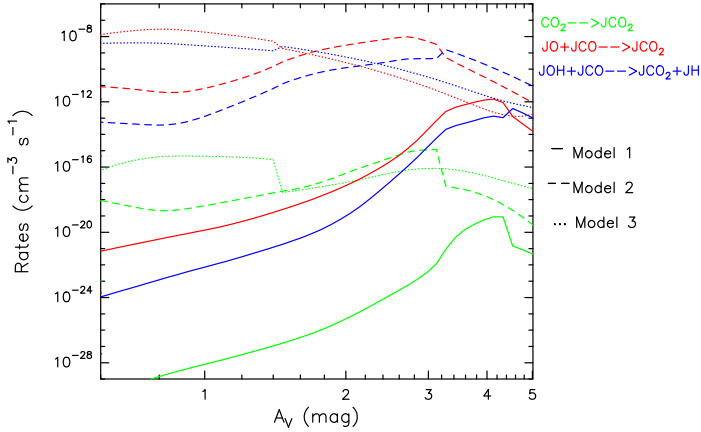}
   \caption{Rates for surface reactions forming CO$_2$ ice. JX means solid X.}          
   \label{figure:CO2_ice_formation_rates}
   \end{figure}

\subsubsection{CO$_2$ ice}
\label{CO_ice}

The rates of the main chemical reactions forming CO$_2$ on dust grains are shown in Fig. \ref{figure:CO2_ice_formation_rates}. 
We obtain that the surface reaction between solid O and solid CO dominates the formation of solid CO$_2$ before reaching its maximum number of monolayers (at $A$$_{\mathrm{V}}$=4.5, 3.5, and 1.5 mag for Models 1, 2, and 3, respectively). For larger extinctions, the CO$_2$ formation becomes dominated by the reaction between solid OH and solid CO. We also find that CO$_2$ depletion barely contributes to the formation of CO$_2$ on dust grains at $A$$_{\mathrm{V}}$$\lesssim$5 mag.

\subsection{Desorption probabilities}
\label{Comparison_delta_ice}

\begin{figure}
\centering
\includegraphics[scale=0.39, angle=0]{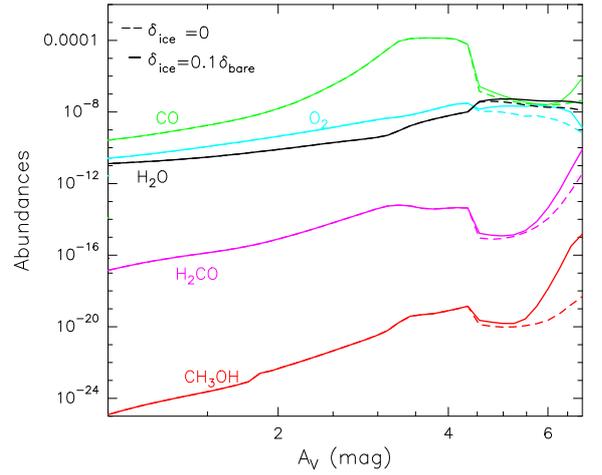} \vspace{0.5cm} \\
\caption{Comparison of the gas-phase fractional abundances, $n$(x)/$n$$_{\mathrm{H}}$, of CO, O$_2$, H$_2$O, H$_2$CO, and CH$_3$OH
for Model 1 ($G$$_{0}$=10$^4$ $\&$ $n$$_{\mathrm{H}}$=10$^4$ cm$^{-3}$), considering $\delta$$_{\mathrm{ice}}$=0 (dashed line) and $\delta$$_{\mathrm{ice}}$=0.1$\delta$$_{\mathrm{bare}}$ (solid line).
}
\label{figure:comparison_delta_ices}
\end{figure}

Table \ref{table:two-body-reactions} in Appendix \ref{Tables} lists the desorption coefficients considering bare grains ($\delta$$_{\mathrm{bare}}$) and icy grains  ($\delta$$_{\mathrm{ice}}$) for each two-body surface reaction. As previously mentioned in Sect. \ref{Reactions_on_grains}, the $\delta$$_{\mathrm{bare}}$ coefficients were obtained theoretically (see expression (2) in Minissale et al. 2016), while the $\delta$$_{\mathrm{ice}}$ coefficients were obtained considering the 10$\%$ of $\delta$$_{\mathrm{bare}}$. In order to analyse the effect of varying this desorption coefficient on the gas-phase abundances of different species, we also considered the case where $\delta$$_{\mathrm{ice}}$=0 to compare it with the case where $\delta$$_{\mathrm{ice}}$=0.1$\delta$$_{\mathrm{bare}}$. The results are shown in Fig. \ref{figure:comparison_delta_ices}. We observe that the difference between not considering desorption when the two-body reaction takes place on icy grains and to consider a small percentage with respect to the desorption when the grains are bare becomes significant at $A$$_{\mathrm{V}}$$\gtrsim$4.5 mag. In particular, this variation in $\delta$$_{\mathrm{ice}}$ implies small differences in the gas-phase abundances of several molecules, such as O$_2$ and H$_2$O, but differences of up to three orders of magnitude for other species, such as CH$_3$OH. We also find that these differences can be larger as the visual extinction increases, demonstrating the importance of considering desorption processes not only with bare substrate, but also with icy grains.

\subsection{Bare versus icy grains}
\label{Comparison_bare_icy_grains}

\begin{figure*}
\centering
\includegraphics[scale=0.41, angle=0]{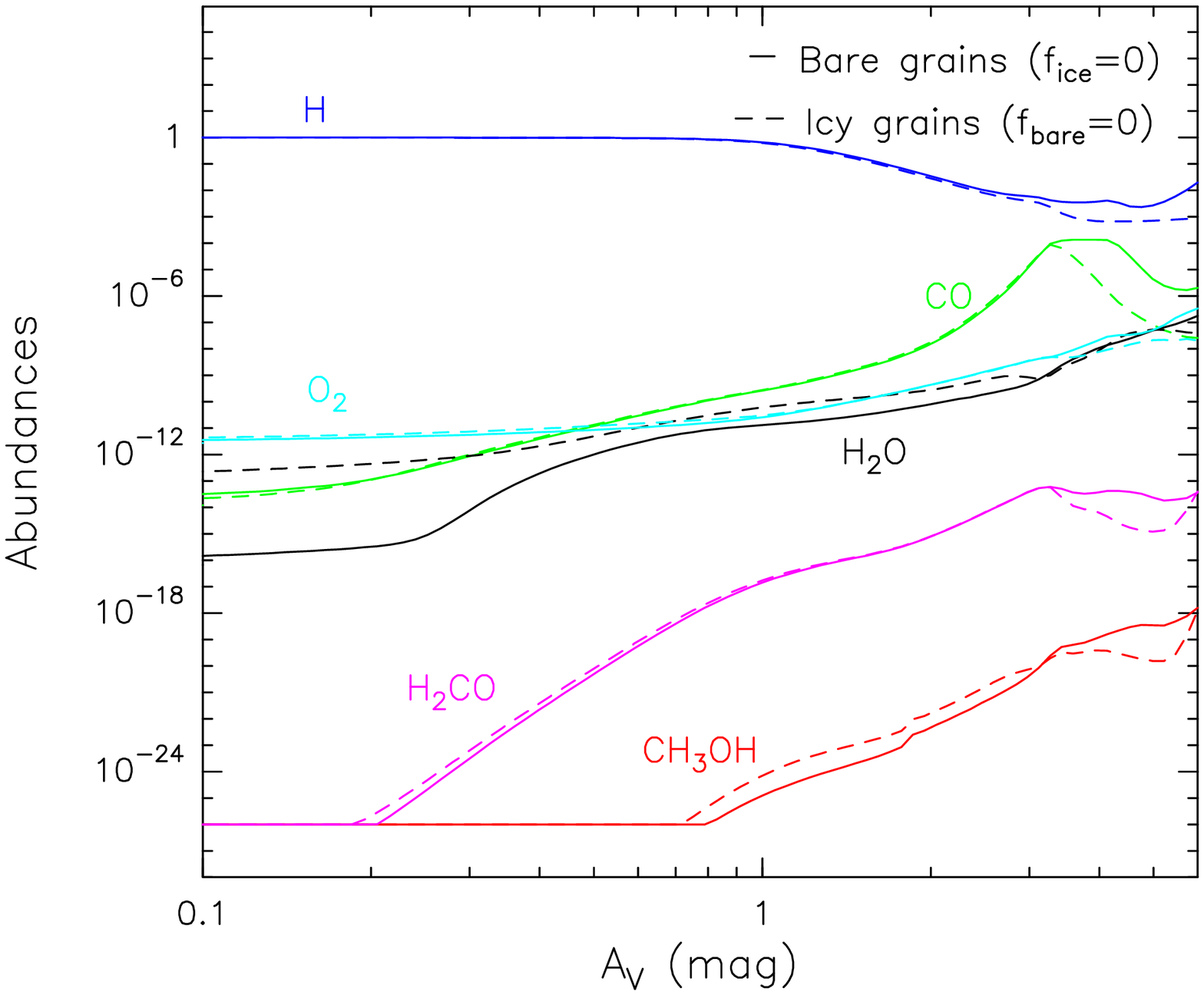}  \hspace{1cm}
\includegraphics[scale=0.41, angle=0]{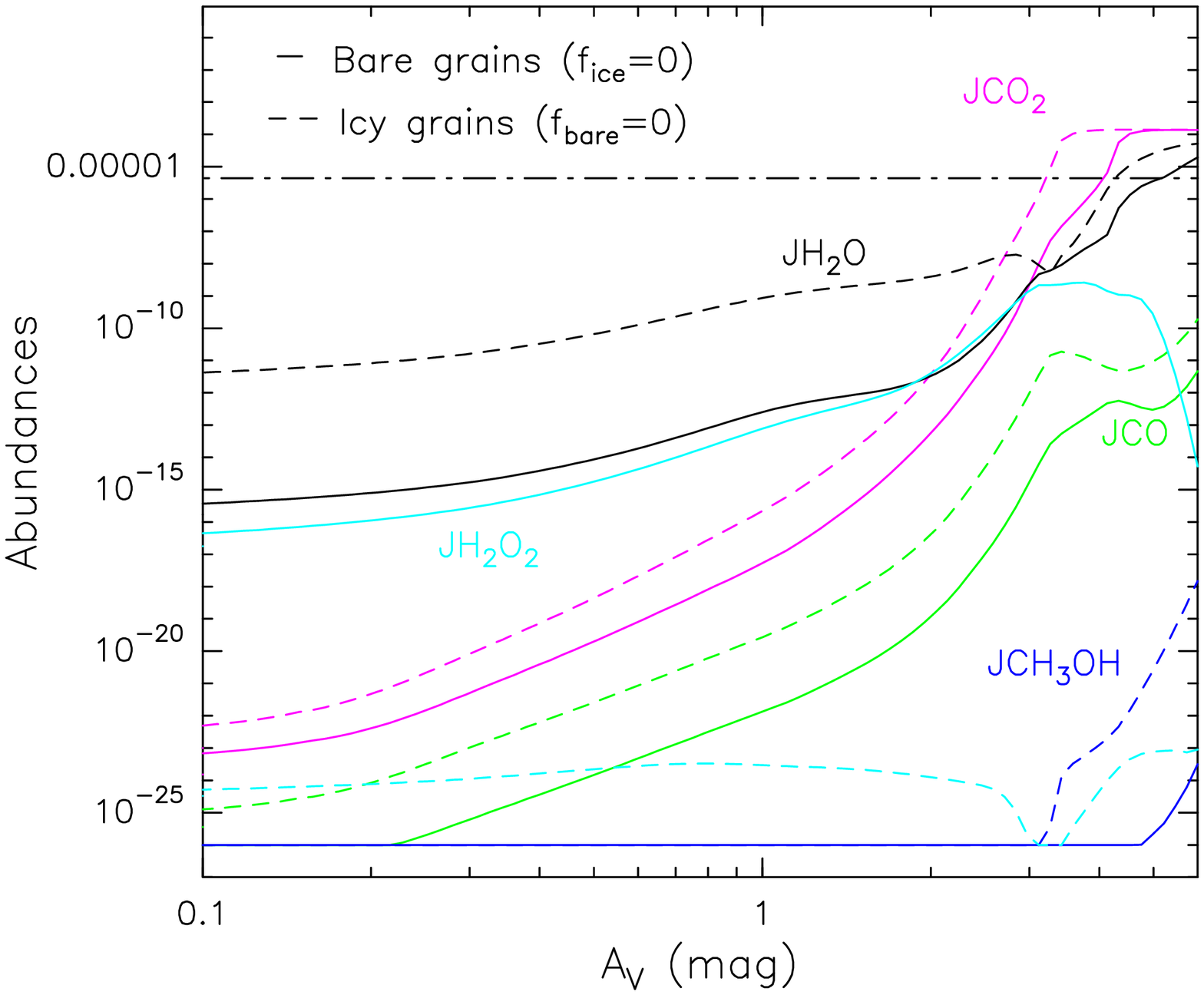}  \hspace{0.7cm}\\
\caption{Comparison of the fractional abundances, $n$(x)/$n$$_{\mathrm{H}}$, for Model 1 ($G$$_{0}$=10$^4$ $\&$ $n$$_{\mathrm{H}}$=10$^4$ cm$^{-3}$), considering bare grains $f$$_{\mathrm{ice}}$=0 (solid line) and icy grains $f$$_{\mathrm{bare}}$=0 (dashed line). Left panel: H, CO, O$_2$, H$_2$O, H$_2$CO, and CH$_3$OH. Right panel: solid CO, CO$_2$, H$_2$O, H$_2$O$_2$, and CH$_3$OH. The dash-dotted black line represents the number of possible attachable sites on grain surfaces per cm$^{3}$ (the limit to form one full monolayer of ice).}
\label{figure:comparison_bare_icy_grains}
\end{figure*}

Although bare grains represent an ideal place for surface chemical reactions to occur, the presence of ice mantles can significantly enrich the gas phase when these ices are desorbed either by thermal or non-thermal processes. In Model 1 ($G$$_0$=10$^4$ and $n$$_{\mathrm{H}}$=10$^4$ cm$^{-3}$), we considered the case in which all grains are bare ($f$$_{\mathrm{ice}}$=0) and the case in which all grains are icy ($f$$_{\mathrm{bare}}$=0) in order to study in more detail the role of ice mantles on dust grains and to quantify the impact on the gas-phase and dust-phase abundances. A comparison of both results is shown in Fig. \ref{figure:comparison_bare_icy_grains}.   
For gas-phase water, unlike other species such as H, O$_2$, and CO, the presence of ice mantles leads to significant differences in its abundance at low extinctions ($A$$_{\mathrm{V}}$$\textless$1 mag). In particular, the water abundance increases up to about three orders of magnitude when grains are icy. This could be due to efficient thermal desorption at these low extinctions, since the water ice reservoir on grains is also significantly larger with the presence of ice mantles for $A$$_{\mathrm{V}}$$<$3 mag. For other species, such as CO, H$_2$CO, and CH$_3$OH, the differences between both cases are significant only for $A$$_{\mathrm{V}}$$\textgreater$3 mag, with a decrease in their gas-phase abundances of up to about two orders of magnitude when grains are icy at 3$<$$A$$_{\mathrm{V}}$$<$6 mag.

Regarding the abundances of solid species (right panel of Fig \ref{figure:comparison_bare_icy_grains}), we obtain large differences for several species at any extinction when considering that grains are bare or icy because of the different binding energies for each case. In addition, we also observe that the presence of ice mantles not only determines the abundance of each species, but also the visual extinction at which a full monolayer of ice is formed. In the case of CO$_2$ and H$_2$O, the formation of the first full monolayer occurs at $\sim$1 mag lower when surface chemistry takes place on icy grains rather than on bare grains. With respect to the abundances, the largest differences are found for solid H$_2$O$_2$, with bare grains promoting its formation. CO, CO$_2$, and H$_2$O also present significant differences between both cases, although these differences become smaller as visual extinction increases, especially for CO$_2$ and water. When all grains are icy, the abundances of solid methanol also increase by about six orders of magnitude at $A$$_{\mathrm{V}}$$<$6 mag.

\subsection{Effect of dust on the chemical composition of PDRs}
\label{dust-chemistry_effect}

\begin{figure}[h!]
\centering
\includegraphics[scale=0.38, angle=0]{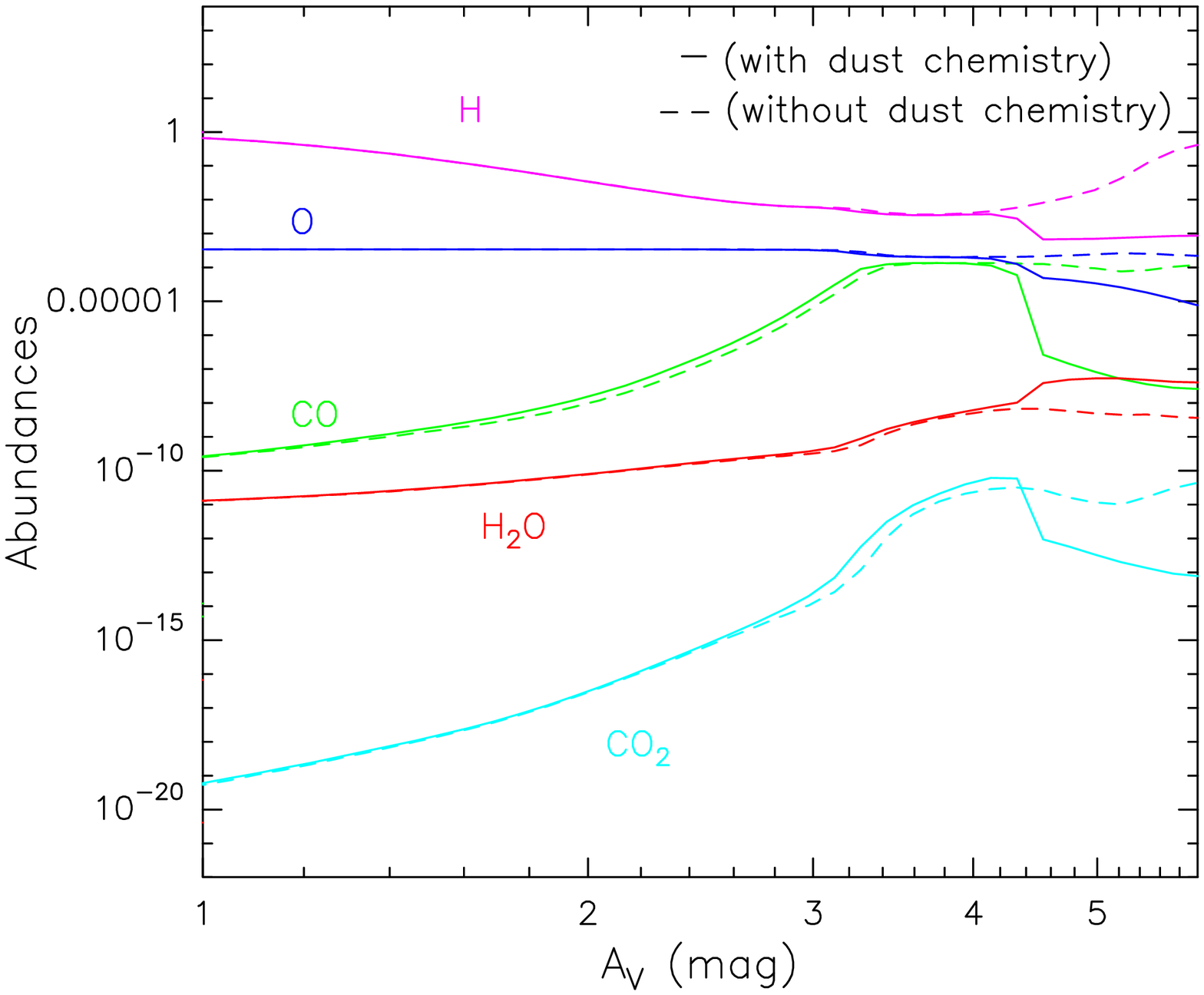} \vspace{0.5cm} \\
\vspace{0.5cm}
\includegraphics[scale=0.38, angle=0]{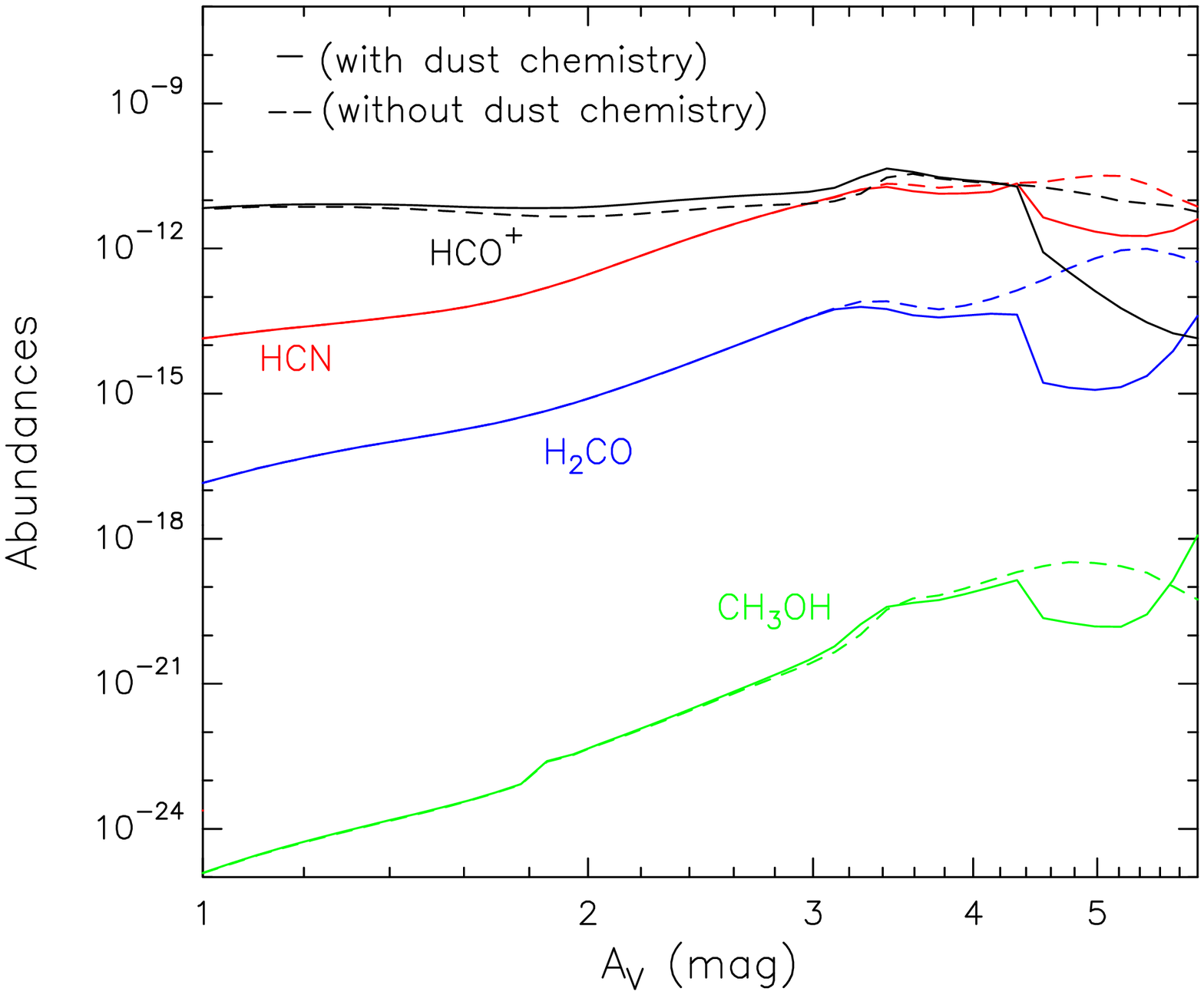}\\
\caption{Comparison of the fractional abundances, $n$(x)/$n$$_{\mathrm{H}}$, for Model 1 ( $G$$_0$=10$^4$ and $n$$_{\mathrm{H}}$=10$^4$ cm$^{-3}$) with and without considering dust chemistry. Top panel: CO, O$_2$, H$_2$O, H, and CO$_2$. Middle panel: CH$_3$OH, H$_2$CO, HCN, and HCO$^+$. 
}
\label{figure:comparison_dust_nodust}
\end{figure}

  \begin{figure}[h!]
   \centering
\includegraphics[angle=0,width=8cm]{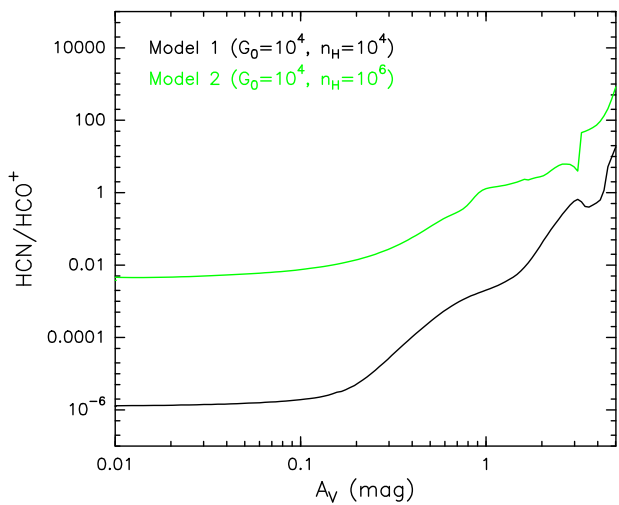}
   \caption{Gas-phase HCN/HCO$^+$ ratio for Models 1 and 2.}          
   \label{figure:ratio_HCN-HCO+}
   \end{figure}

We ran Model 1 (characteristic of starbursts) again considering that H$_2$ formation is the only reaction taking place on dust grains to analyse how the implementation of dust chemistry affects the abundances of gas-phase species. The results for different gas-phase species are shown in Fig. \ref{figure:comparison_dust_nodust}. In the top panel, we present results for the gas-phase abundances of H, O, CO, CO$_2$, and H$_2$O. For low visual extinctions, we barely find differences in the abundances of most of these species, independently of whether we consider  dust chemistry or not. For $A$$_{\mathrm{V}}$$\gtrsim$4 mag, however, we clearly observe a decrease (up to three orders of magnitude) in all their abundances as a result of the formation of ices, when adsorption, desorption, and two-body processes on grain surfaces, along with the incidence of UV photons and cosmic rays on dust grains, are considered in the chemical network.

The bottom panel of Fig. \ref{figure:comparison_dust_nodust} shows the abundances as a function of $A$$_{\mathrm{V}}$ for H$_2$CO and CH$_3$OH in the gas phase. 
We again observe large differences between both cases at $A$$_{\mathrm{V}}$$\gtrsim$4 mag. In particular, at $A$$_{\mathrm{V}}$$\sim$5 mag, we observe a change in the trend of H$_2$CO and CH$_3$OH gas-phase abundances; while their abundances start to decrease in the case without dust chemistry, they start to increase by several orders of magnitude when dust chemistry is implemented in the code, since they are mainly formed on grain surfaces (Chutjian et al. 2009). A similar trend is found for HCN, revealing the importance of considering grain chemical processes to explain the enhanced abundance of these molecules in the gas phase (Lintott \& Viti 2006; Akimkin et al. 2013).  
In Fig. \ref{figure:ratio_HCN-HCO+}, we show the gas-phase HCN/HCO$^{+}$ ratio obtained when surface chemistry is considered in the models with the highest radiation field intensities (Models 1 and 2). Although we find an increasing HCN/HCO$^+$ ratio for $A$$_{\mathrm{V}}$$\gtrsim$0.1 mag in both cases, this increase is particularly pronounced at $A$$_{\mathrm{V}}$$\gtrsim$3.5 mag because of the presence of surface chemistry on grains.

\subsection{Comparison with the original Meijerink PDR code}
\label{Comparison}

\begin{figure*}
\centering
\includegraphics[scale=0.415, angle=0]{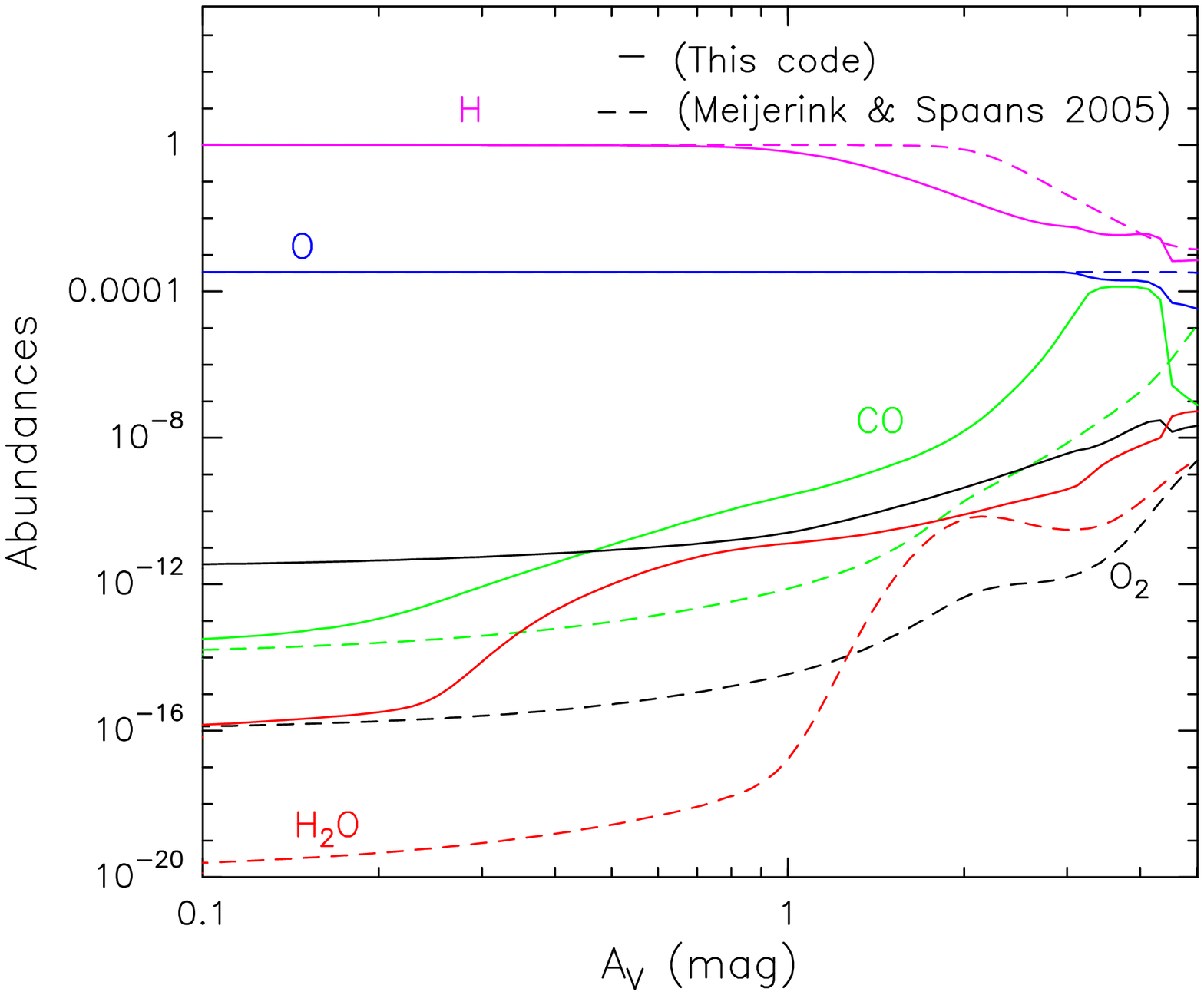}  \hspace{1cm}
\includegraphics[scale=0.415, angle=0]{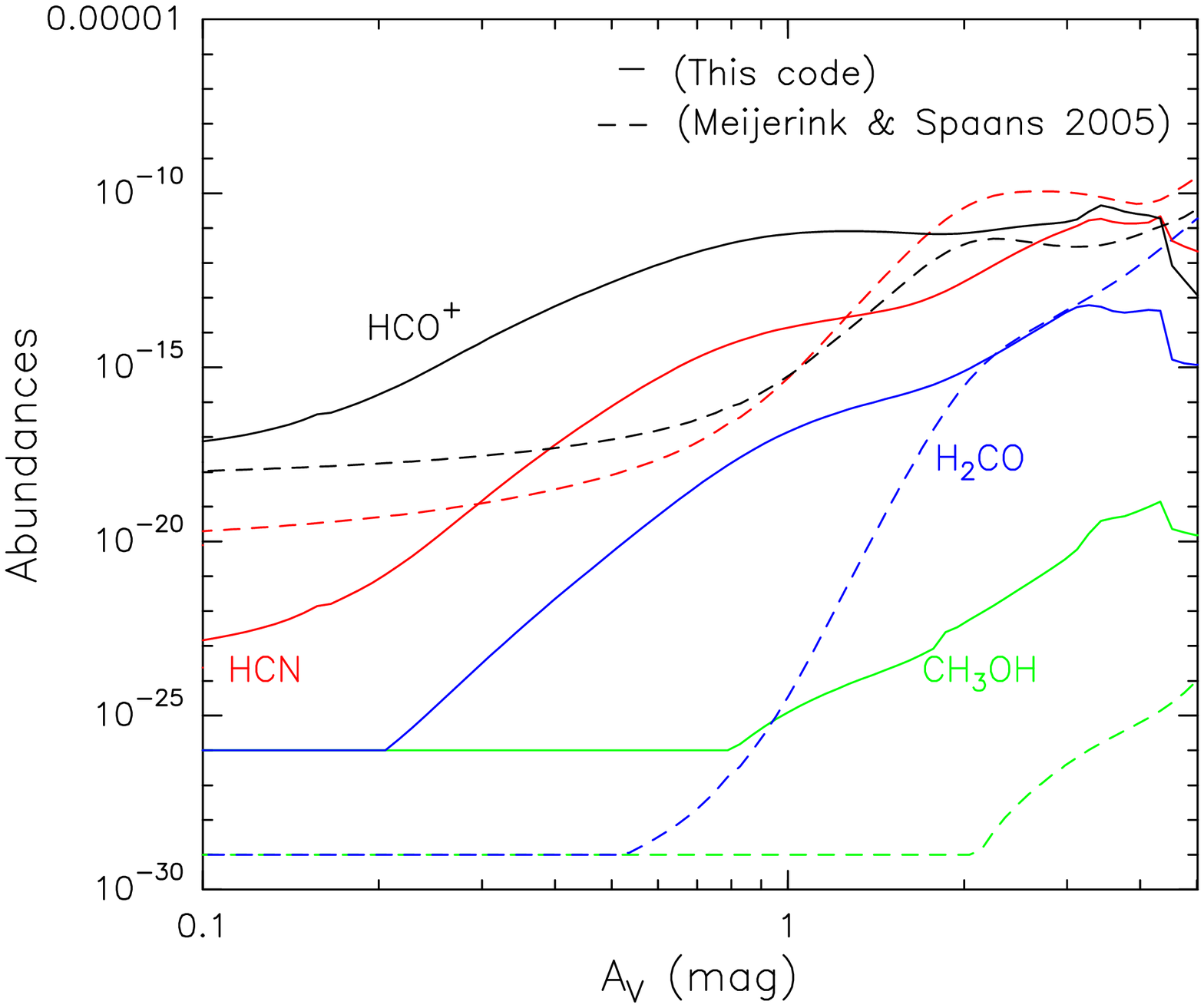}  \hspace{0.7cm}\\
\caption{Comparison of the fractional abundances, $n$(x)/$n$$_{\mathrm{H}}$, for Model 1 ($G$$_{0}$=10$^4$ $\&$ $n$$_{\mathrm{H}}$=10$^4$ cm$^{-3}$), with those obtained from the original version of the {\it{Meijerink}} code (Meijerink \& Spaans 2005). Left panel: H, CO, O, O$_2$, and H$_2$O. Right panel: CH$_3$OH, H$_2$CO, HCN, and HCO$^+$.}
\label{figure:comparison_with_Rowin_123_I}
\end{figure*}

\begin{figure*}
\centering
\includegraphics[scale=0.415, angle=0]{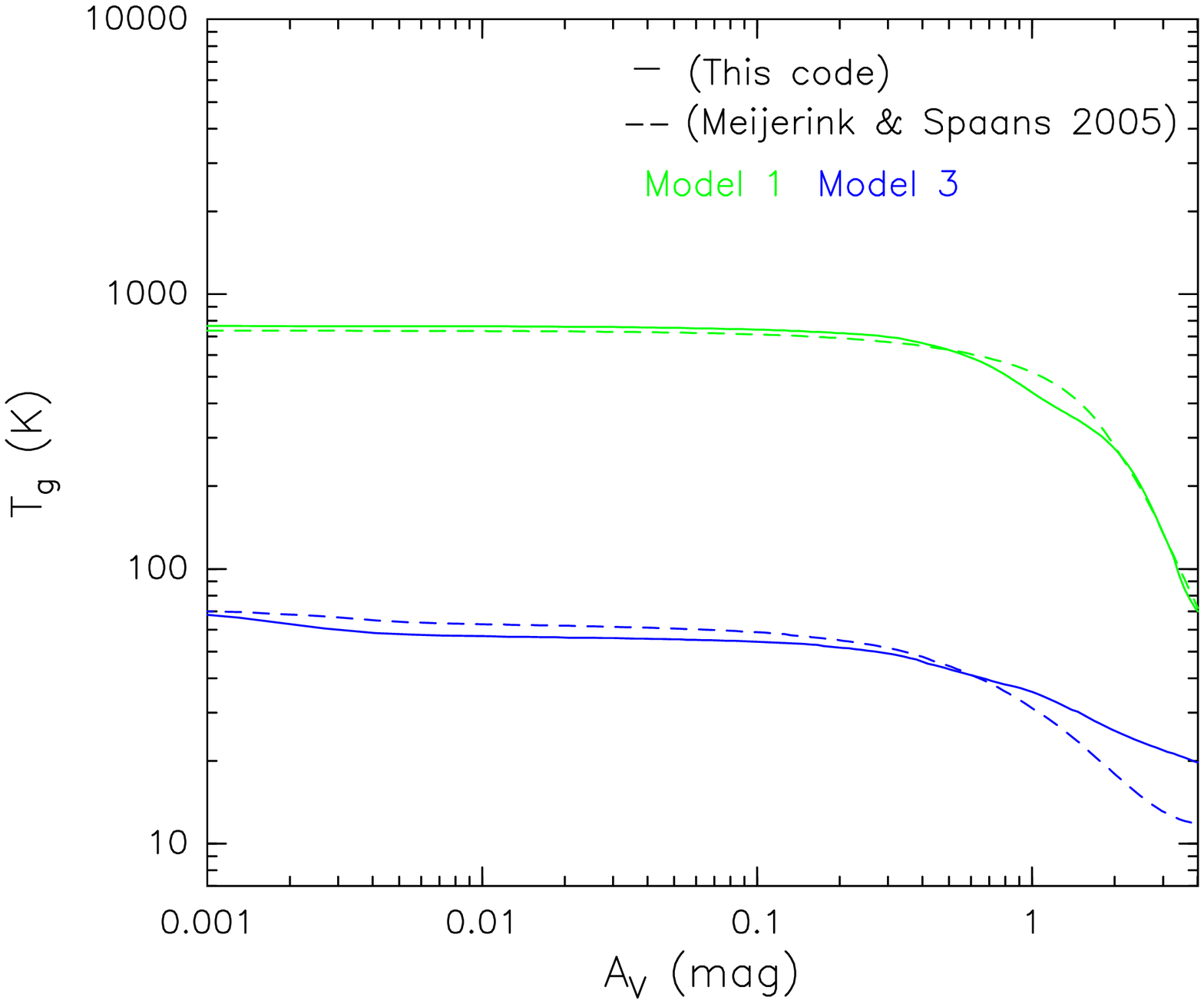}  \hspace{1cm}
\includegraphics[scale=0.415, angle=0]{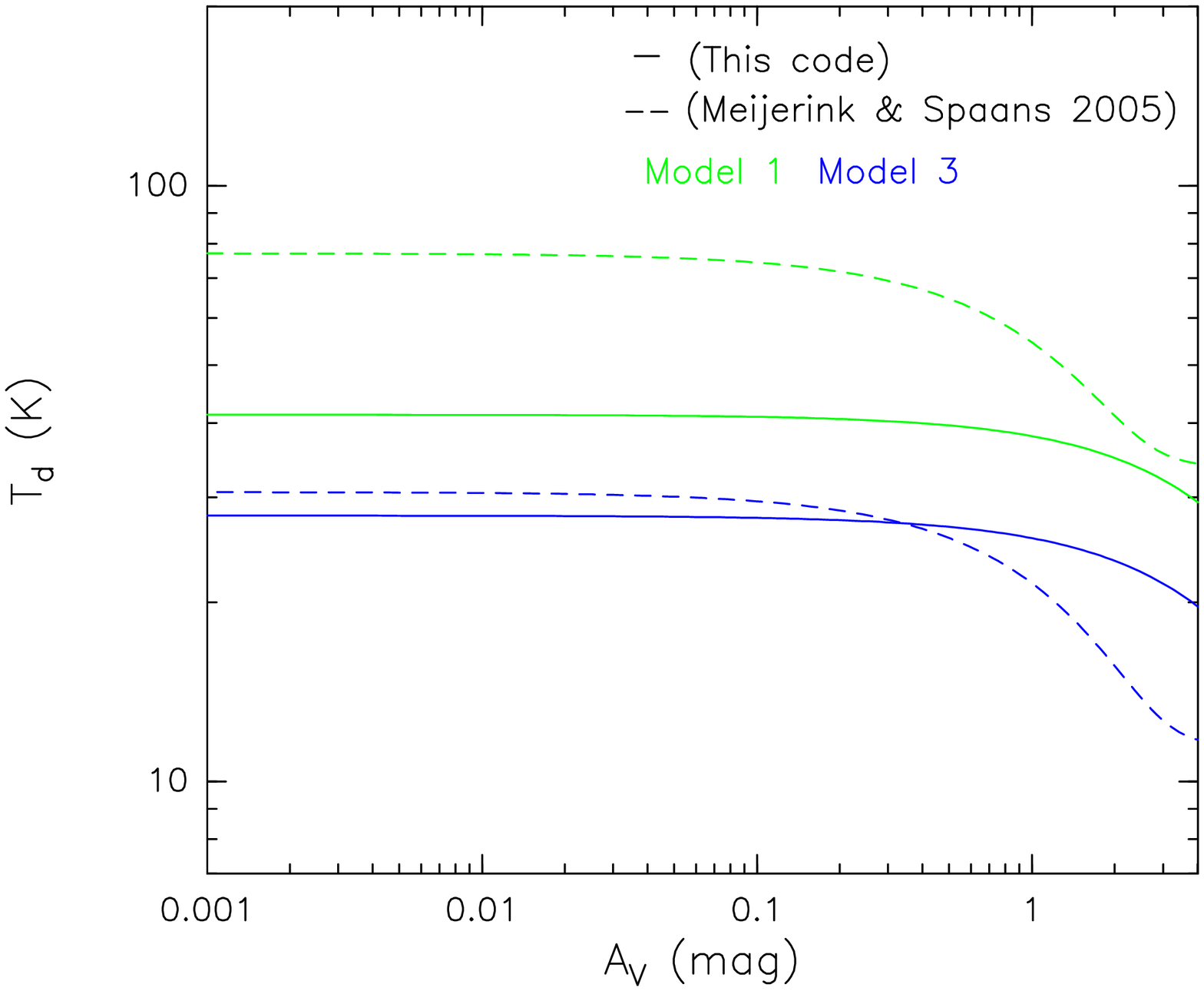}  \hspace{0.7cm}\\
\caption{Comparison of the gas temperature (left) and dust temperature (right) for Models 1 ($G$$_{0}$=10$^4$ $\&$ $n$$_{\mathrm{H}}$=10$^4$ cm$^{-3}$) and 3 ($G$$_{0}$=10$^2$ $\&$ $n$$_{\mathrm{H}}$=10$^6$ cm$^{-3}$) with those obtained from the original version of the {\it{Meijerink}} code (Meijerink \& Spaans 2005).}
\label{figure:comparison_Tgas}
\end{figure*}

We also compared some of our results 
with those obtained by Meijerink \& Spaans (2005) using the original version\footnote{In the original version of the {\it{Meijerink}} code, the gas-phase chemical network was taken from UMIST 1999.} of the {\it{Meijerink}} code, in which H$_2$ formation was the only chemistry considered on dust grains. In Fig. \ref{figure:comparison_with_Rowin_123_I} (left panel), we observe that in the original {\it{Meijerink}} code the decrease of the H abundance to form H$_2$ is slightly sharper than in the current version of the code. At $A$$_{\mathrm{V}}$$<$4 mag, while atomic oxygen does not present significant differences between both models, the update of the gas chemical network is mainly responsible for the increase of the abundances of CO, H$_2$O, and O$_2$ by several orders of magnitude. For higher extinctions, we also obtain large differences between both versions of the $Meijerink$ code (higher abundances for H, O, and CO, and lower for O$_2$ and H$_2$O in the original code), but in this case, these differences are mainly due to the implementation of dust chemistry.
For other molecules, such as HCO$^+$, H$_2$CO, and CH$_3$OH (right panel), we observe the same effect at $A$$_{\mathrm{V}}$$\lesssim$4 mag as that observed in the molecules of the top panel. 
At $A$$_{\mathrm{V}}$$>$4 mag, the differences observed between both versions of the code for HCO$^+$ and CH$_3$OH are due to both the implementation of more than 3000 new gas-phase reactions and to surface chemistry, leading to a decrease and increase, respectively, of their gas-phase abundances of several orders of magnitude.

With respect to the thermal balance, we considered a more recent analytical expression for the dust temperature (obtained from Garrod $\&$Pauli 2011, Sect. \ref{Dust_temperature}) than that considered in the original version of the {\it{Meijerink}} code (from Hollenbach et al. 1991). As we see in Fig. \ref{figure:comparison_Tgas} (right panel), both expressions lead to very distinct values of $T$$_{\mathrm{d}}$, especially for a high $G$$_{0}$ model (Model 1) at $A$$_{\mathrm{V}}$$\textless$0.5 mag with a difference of more than 30 K. For 0.5$<$$A$$_{\mathrm{V}}$$<$5 mag, the differences for the dust temperature between both codes become smaller for the high $G$$_0$ PDR and larger for the the low $G$$_0$ and high density PDR.
Deep in the cloud, different dust temperature expressions significantly affect the gas temperature, determining how its profile decreases as the visual extinction increases. In the left panel of Fig. \ref{figure:comparison_Tgas}, we obtain differences of a few hundreds degrees between the two cases at $A$$_{\mathrm{V}}$$\sim$1 mag in Model 1, and of a few tens of degrees at $A$$_{\mathrm{V}}$$\gtrsim$2 mag in Model 3.
These differences are mainly due to the influence of $T$$_{\mathrm{d}}$ on $T$$_{\mathrm{g}}$ through the heating and cooling rates. Variations in the reaction rates (for example, produced when considering a different chemical network with updated parameters for each reaction) can originate large differences in the number density of species with a direct impact on several heating and cooling processes (e.g. molecular cooling) and, therefore, in the final gas temperature of each model as well.

\section{Comparison to observations}
\label{Comparison_observations}

\subsection{The Horsehead case}
\label{Horsehead}

The Horsehead nebula photodissociation region ($d$$\sim$400 pc) is located at the western edge of the molecular cloud L1630 illuminated by the O9.5V star $\sigma$ Ori (Habart et al. 2005). The far-UV intensity of the incident radiation field illuminating the Horsehead nebula is $\chi$$\sim$60 in Draine{\footnote{Draine field$\simeq$1.7$\times$Habing field.}} units ($G$$_0$$\sim$100; Habart et al. 2005, Goicoechea et al. 2009), which is moderate compared to those of classical PDRs illuminated by O stars (generally $\chi$$\sim$10$^4$-10$^5$; e.g. Tielens et al. 1993). In the last decade, several molecular observations were carried out in this region to study, for example, sulphur chemistry (Goicoechea et al. 2006), the presence of ions, such as  C$^+$ and CF$^+$ (Guzm\'an et al. 2012), and the presence of organic molecules, such as CCH, C$_4$H, H$_2$CO, and CH$_3$OH (Goicoechea et al. 2009, Guzm\'an et al. 2011, 2013). In particular, recent results from a theoretical analysis of the chemistry forming H$_2$CO and CH$_3$OH indicate that the observations of these molecules in the PDR position of the Horsehead cannot be reproduced when only gas-phase chemistry is considered in the model (Guzm\'an et al. 2013).

\begin{figure*}
\centering
\includegraphics[scale=0.375, angle=0]{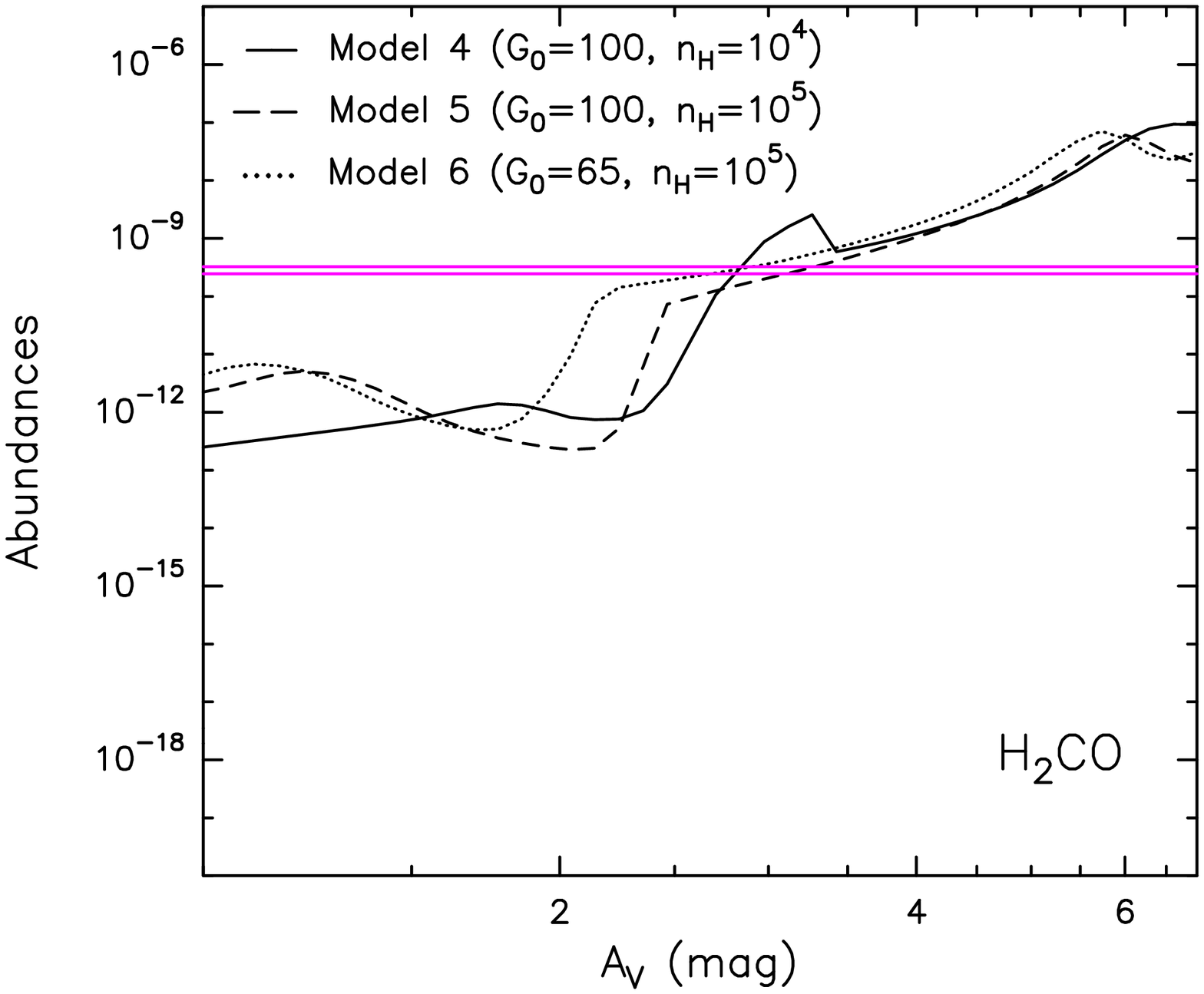}  \hspace{1cm}
\includegraphics[scale=0.375, angle=0]{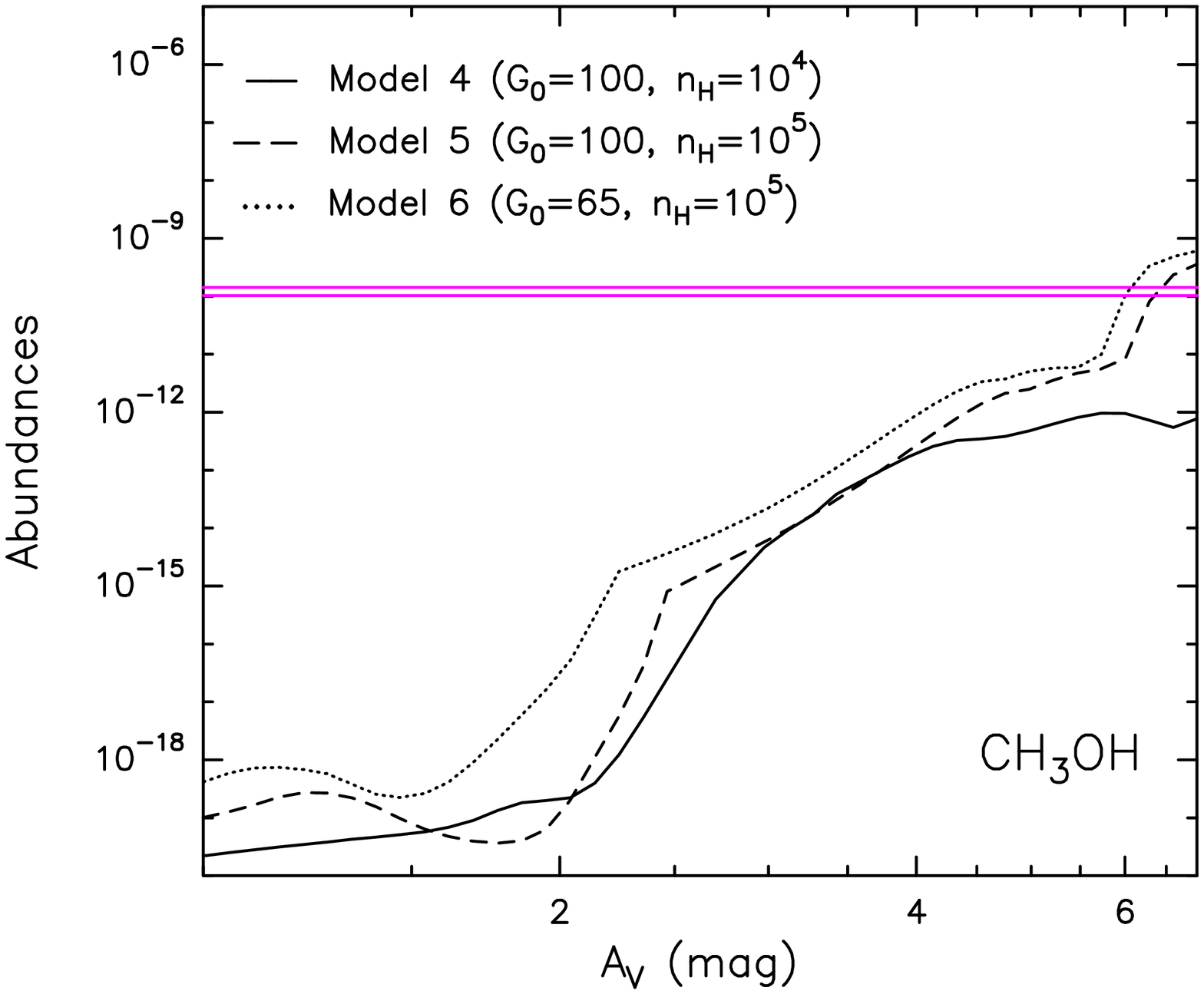}  \hspace{0.7cm}\\
\caption{Comparison of the H$_2$CO (left) and CH$_3$OH (right) abundances, $n$(x)/$n$$_{\mathrm{H}}$, from Models 4, 5, and 6 with observations from the Horsehead PDR. The range of observational abundances for each molecule is shown with magenta lines.}
\label{figure:observations_Horsehead}
\end{figure*}

We considered models with a radiation field intensity of $G$$_0$=100 (Models 4 and 5 with densities of 10$^4$ and 10$^5$ cm$^{-3}$, respectively) and $G$$_0$=65 (Model 6) to reproduce the conditions of the Horsehead PDR to analyse the origin of H$_2$CO and CH$_3$OH forming on dust grain surfaces. The results for the abundances of H$_2$CO and CH$_3$OH as a function of the visual extinction are shown in Fig. \ref{figure:observations_Horsehead}. We also included the value range for the observed abundances for both molecules (magenta lines) in the plots. These observations are taken from Guzm\'an et al. (2013). The left panel of Fig. \ref{figure:observations_Horsehead} shows that the observations of H$_2$CO are reproduced by the three models at $A$$_{\mathrm{V}}$$\sim$2.5-3 mag and the right panel of Fig. \ref{figure:observations_Horsehead} shows the model results for CH$_3$OH. In this case, we observe that only models with high density ($n$=10$^5$ cm$^{-3}$, Models 5 and 6) reproduce the observations at $A$$_{\mathrm{V}}$$\sim$6 mag. We also obtain that the difference in the radiation field intensity mainly determines the visual extinction at which the observations are reproduced (with a difference of $\sim$0.5 mag). 

\begin{figure*}
\centering
\includegraphics[scale=0.375, angle=0]{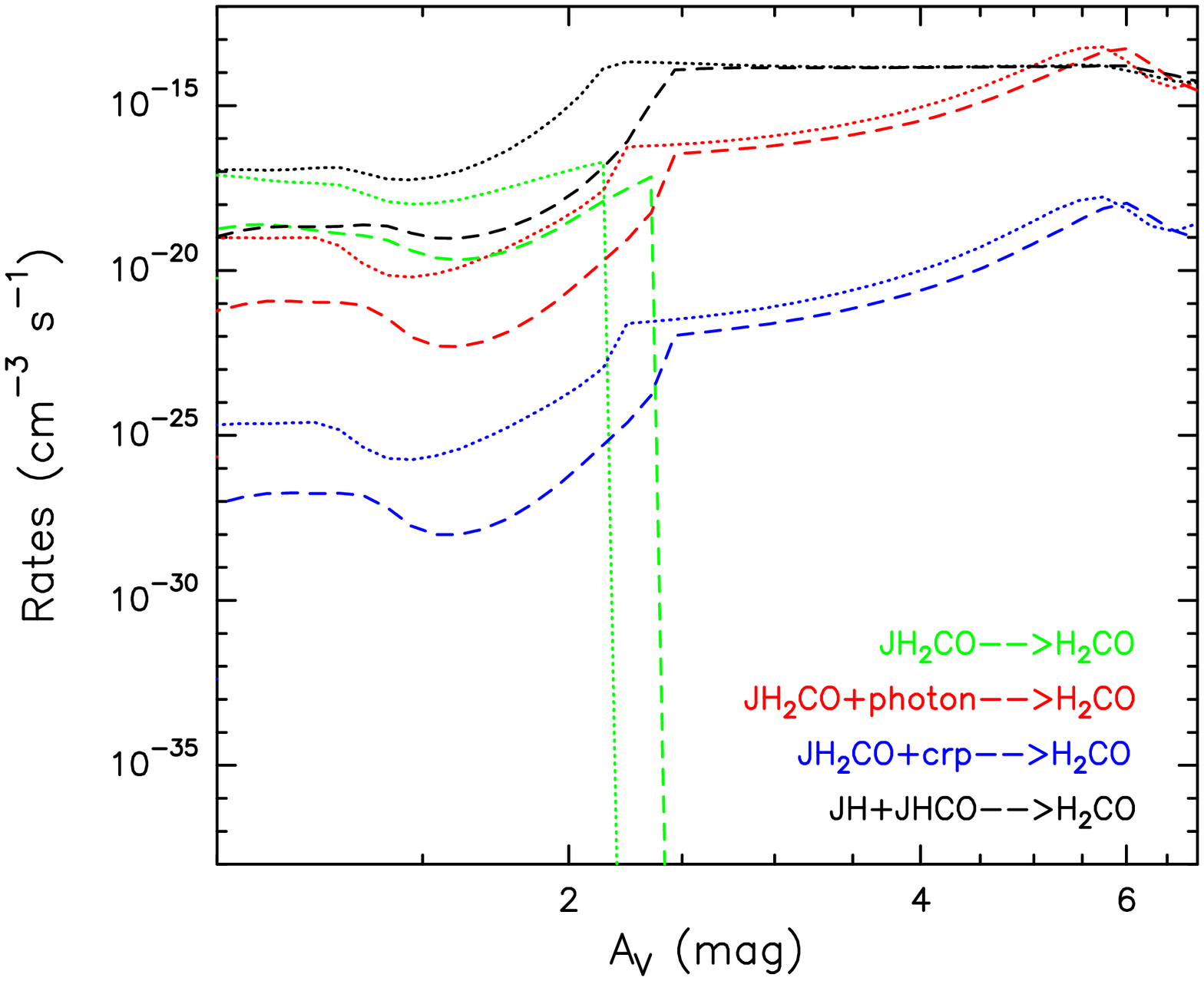}  \hspace{1cm}
\includegraphics[scale=0.375, angle=0]{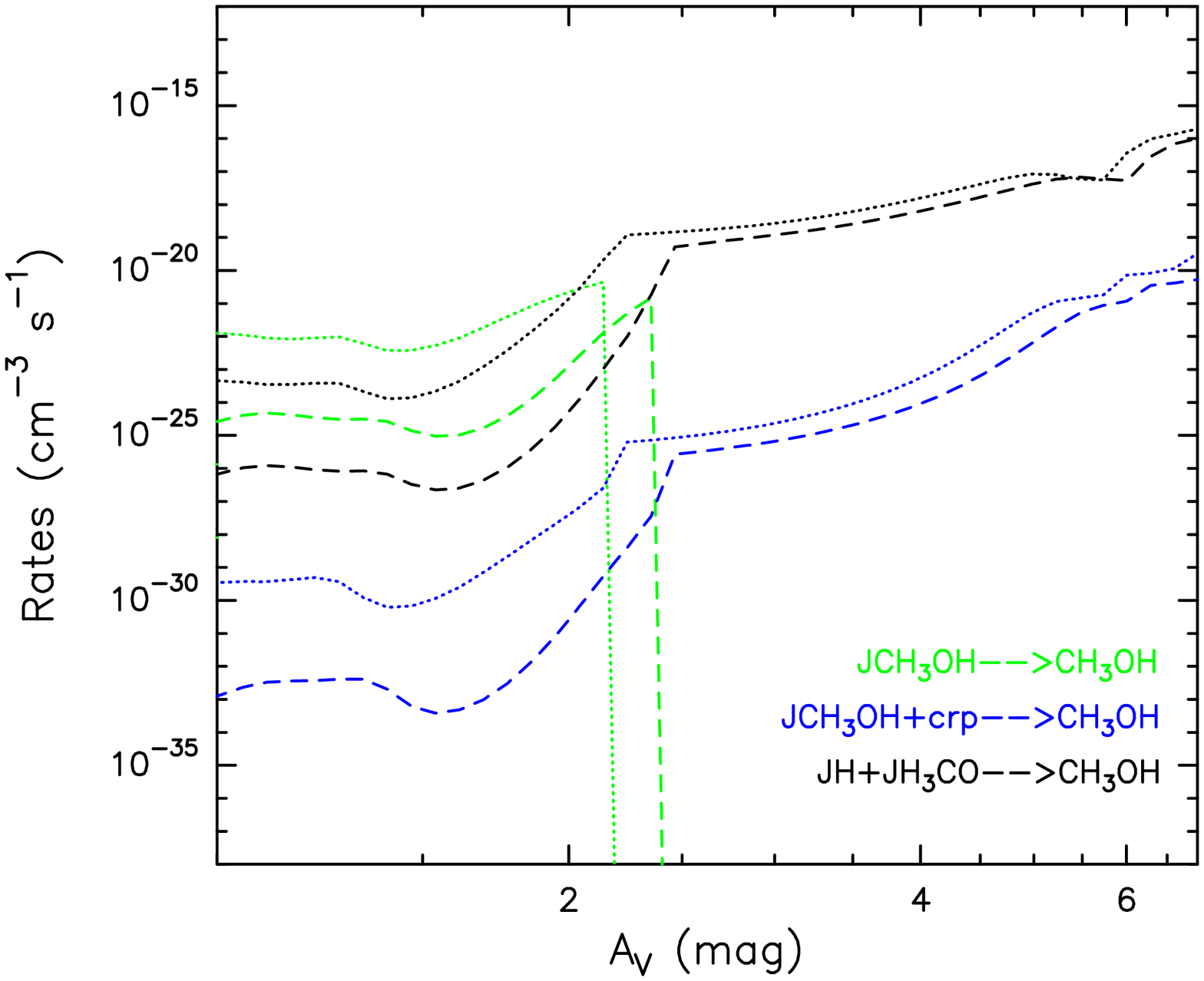} \hspace{0.7cm}\\
\caption{Surface reaction rates forming H$_2$CO (left) and CH$_3$OH (right) gas for Models 5 (dashed line) and 6 (dotted line). JX means solid X.}
\label{figure:formation_rates_1136-145}
\end{figure*}

Figure \ref{figure:formation_rates_1136-145} shows the rates of the surface reactions leading to H$_2$CO (left panel) and CH$_3$OH (right panel) gas that we considered in our code. The results are for Models 5 and 6; those reproducing the abundances of both molecules, H$_2$CO and CH$_3$OH. In particular, for the case of H$_2$CO, we considered thermal desorption, direct cosmic-ray and photodesorption, and formation of H$_2$CO gas through chemical desorption coming from the reaction between solid hydrogen and solid HCO. For the CH$_3$OH formation, we also considered thermal desorption, cosmic-ray desorption, and the surface reaction between solid hydrogen and solid H$_3$CO. We did not include, however, direct photodesorption, since laboratory experiments from Bertin et al. (2016) show its minor role in the formation of gas methanol, as mentioned in Sect. \ref{Photo-processes_on_grains}. In both models we obtain that H$_2$CO gas is mainly formed through the surface reaction between solid hydrogen and solid HCO, whose rates are more than two orders of magnitude higher than those for direct photodesorption at $A$$_{\mathrm{V}}$$\lesssim$4.5 mag. We find that direct photodesorption has a high contribution in the formation of H$_2$CO gas only at $A$$_{\mathrm{V}}$$>$4.5 mag. For the same models (5 and 6), we also obtain that methanol is mainly formed at $A$$_{\mathrm{V}}$$\sim$6 mag through chemical desorption upon the surface reaction between solid H and solid H$_3$CO. These findings disagree with recent results from Guzm\'an et al. (2013), who concluded that direct photodesorption is the main process of releasing methanol into the gas phase where thermal desorption of ice mantles is ineffective. This discrepancy is due to the overestimation of their photodesorption rates (Bertin et al. 2016).

\begin{figure*}
\centering
\includegraphics[scale=0.375, angle=0]{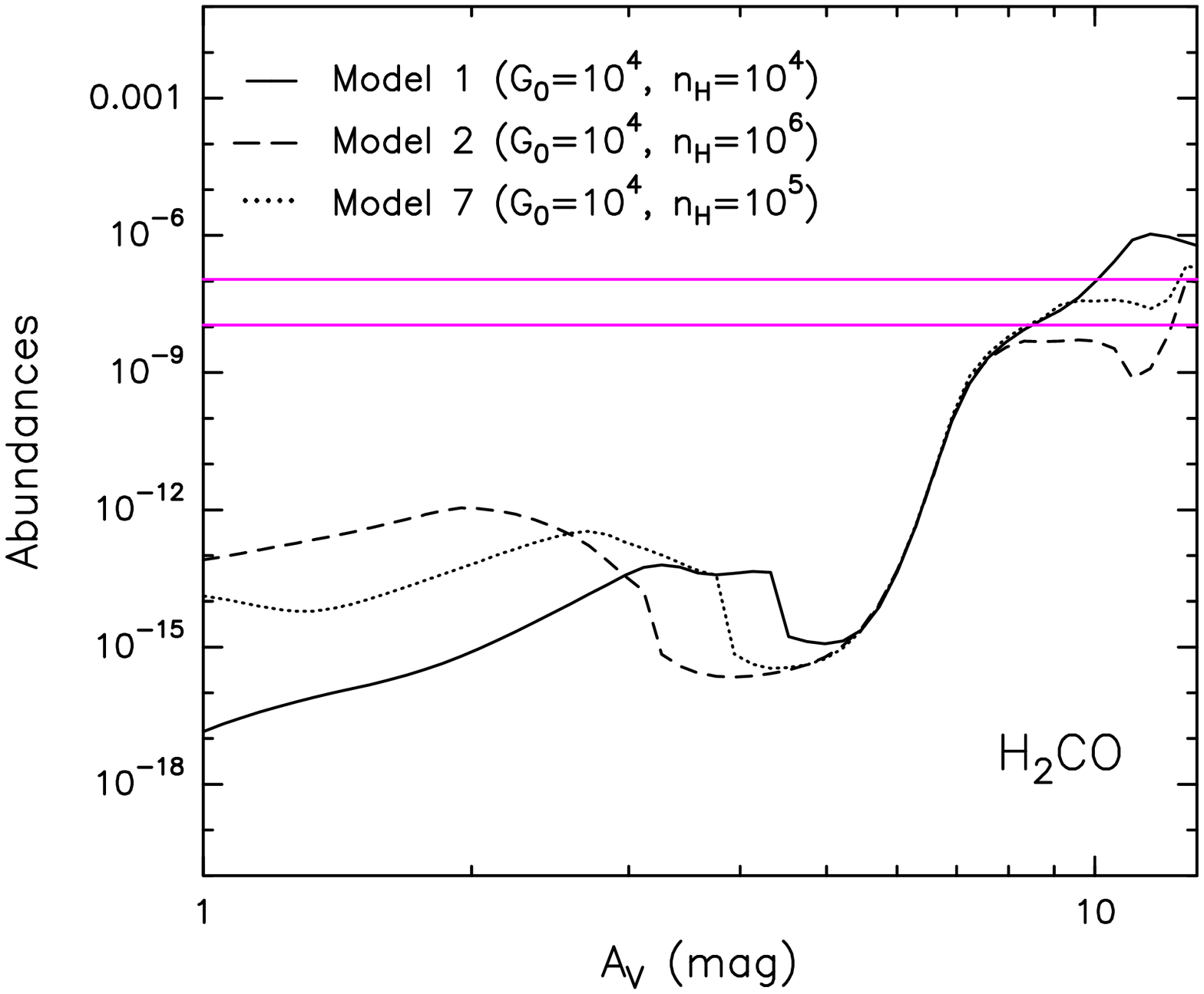}  \hspace{1cm}
\includegraphics[scale=0.375, angle=0]{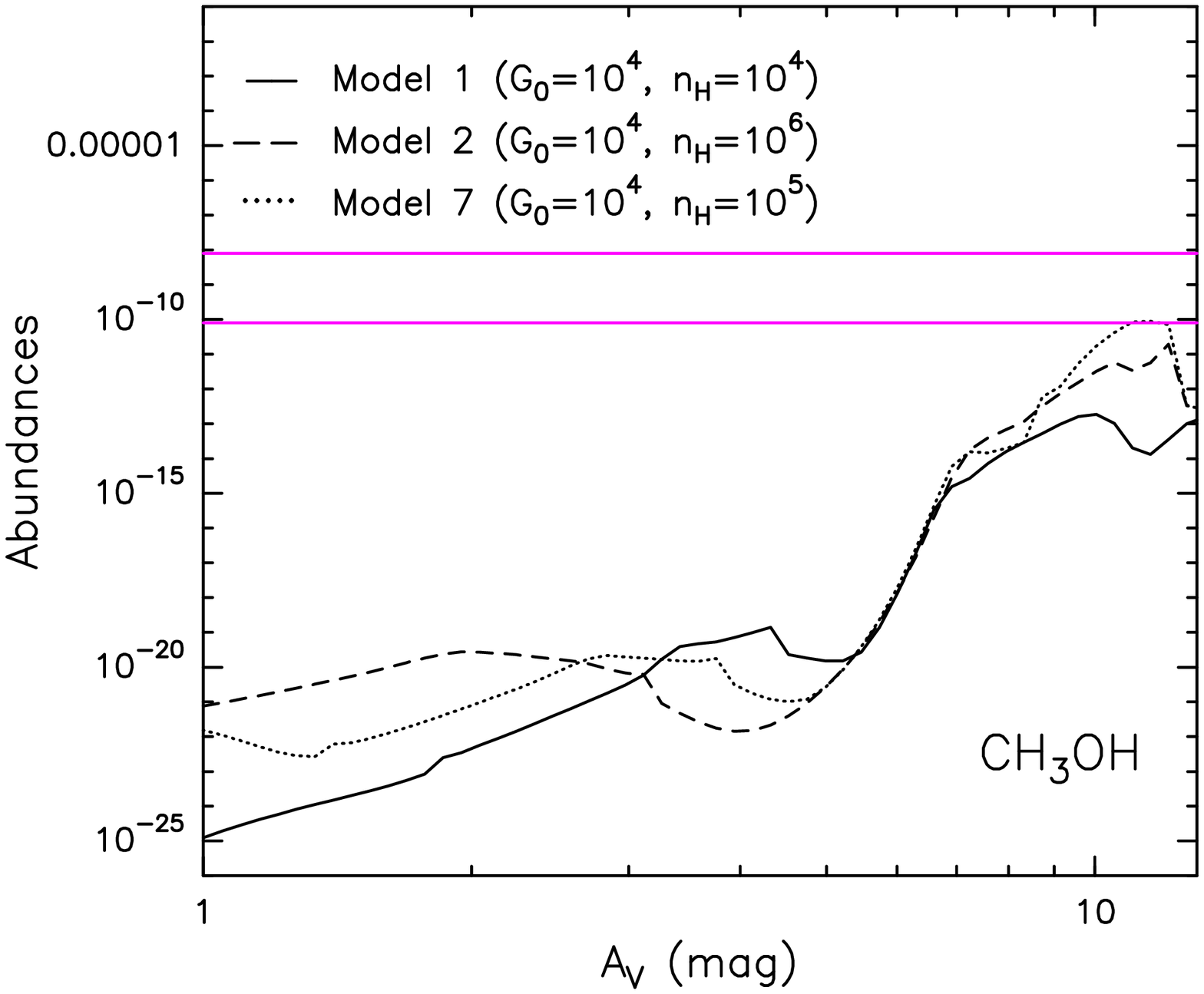}  \hspace{0.7cm}\\
\caption{Comparison of the H$_2$CO (left) and CH$_3$OH (right) abundances, $n$(x)/$n$$_{\mathrm{H}}$, from Models 1, 2, and 7 with observations from the Orion Bar PDR. The range of observational abundances for each molecule is shown with magenta lines.}
\label{figure:observations}
\end{figure*}


\subsection{The Orion Bar case}
\label{Observations}

The Orion Bar ($d$$\sim$414 pc) is located between the Trapezium cluster and the Orion molecular cloud (Menten et al. 2007). It is the prototypical warm PDR with a FUV radiation field at the ionisation front of $G$$_0$$\sim$(1-4)$\times$10$^4$ (Marconi et al. 1998) and densities of $n$$\sim$10$^4$-10$^6$ cm$^{-3}$ (Leurini et al. 2010).

We also analyse the origin of H$_2$CO and CH$_3$OH gas in the interclump of the Orion Bar. For that, we considered observations from Leurini et al. (2010) and Models 1 and 2, as well as a new model with $G$$_0$=10$^4$ and $n$$\sim$10$^5$ cm$^{-3}$. Figure \ref{figure:observations} shows the model results for the H$_2$CO and CH$_3$OH gas abundances and the range of the observational abundances (magenta lines). We note that observations of H$_2$CO and CH$_3$OH are only reproduced at $A$$_{\mathrm{V}}$$>$7.5 mag and $>$10 mag, respectively. Since our code is a steady-state chemical code, we would need the implementation of time dependence to study the chemistry occurring in this visual extinction range properly.


\section{Summary and conclusions}
\label{conclusions}

We have updated the {\it{Meijerink}} PDR code by implementing 3050 new gas-phase reactions and surface chemistry. In particular, we include 117 surface reactions including adsorption, thermal and non-thermal desorption, two-body reactions, photo processes and cosmic-ray processes on dust grains. In total, we used 7621 chemical reactions in this new version of the {\it{Meijerink}} code.  In our dust treatment, we also distinguish between icy and bare grains. 
We analysed in detail how the variation of density and the intensity of the radiation field influence the thermal balance and gas/surface chemistry. In addition, we also studied how chemistry is affected by the presence/absence of ice mantles on dust grains. Our code also gives an indication of the origin of the formation of ices, such as CO$_2$ and water, that are present in different environments and, by comparing with observations, it shows the most efficient surface processes forming H$_2$CO and CH$_3$OH gas. The main results can be summarised as follows:

\begin{enumerate}
\renewcommand{\labelenumi}{$\bullet$ }
\item {\it Ice formation}: The formation of a large number of H$_2$O ice monolayers is favoured by high densities.
Varying the radiation field intensity mainly determines the threshold extinction to form ices of H$_2$O and, especially, of CO$_2$ (ice formation at lower $A$$_{\mathrm{V}}$ for low $G$$_0$). Our models also indicate that a low $G$$_0$ promotes the formation of CH$_3$OH and CO ices. In particular:

\vspace{0.5cm}

\begin{enumerate}
\renewcommand{\labelenumi}{$\bullet$ }
\item {\it{H$_2$O ice formation}}: This formation is mainly dominated by the surface reaction between solid H and solid OH at $A$$_{\mathrm{V}}$$\lesssim$5 mag. H$_2$O$_2$ on dust grains is also an important intermediate in the formation of solid water for a high $G$$_0$ PDR.
\vspace{0.2cm}

\item {\it CO$_2$ ice formation}: This formation is mainly dominated by the surface reaction between solid O and solid CO. Once the maximum number of monolayers is reached, the main reaction producing CO$_2$ on dust grains is the reaction between solid OH and solid CO. 
\vspace{0.2cm}

\end{enumerate}

\end{enumerate} 


\begin{enumerate}
\renewcommand{\labelenumi}{$\bullet$ }
\item {\it Bare versus icy grains}: The presence or absence of ice mantles on dust grains determines the (gas and solid) abundances of many species leading to differences of several orders of magnitude for many of them, especially at $A$$_{\mathrm{V}}$$\gtrsim$3 mag. Some of the most affected species are water, methanol, CO, and H$_2$O$_2$ with differences of at least two orders of magnitude. The type of substrate (bare or icy) also varies the extinction at which ices of CO$_2$ and H$_2$O are formed with differences of up to 1 mag.  
\end{enumerate}

\begin{enumerate}
\renewcommand{\labelenumi}{$\bullet$ }
\item {\it Desorption probabilities}: The difference between not considering chemical desorption with icy grains and to consider a small ($10\%$) probability of desorption with respect to that for bare grains leads to variations of up to two-three orders of magnitude in the gas-phase abundances of molecules, such as methanol. 
\end{enumerate}

\begin{enumerate}
\renewcommand{\labelenumi}{$\bullet$ }
\item {\it PDR observations}: The main mechanism releasing H$_2$CO and CH$_3$OH into the gas phase of the Horsehead PDR is chemical desorption upon the surface reaction between solid H with solid HCO and H$_3$CO, respectively, rather than photodesorption.  
\end{enumerate}

These results demonstrate the important role of dust grains when studying chemistry in molecular clouds exposed to different UV radiation fields. They also show the need to implement time dependence in order to properly analyse the chemical composition at large visual extinctions.

\begin{acknowledgements}
This work is supported by the Netherlands Organization for Scientific Research (NWO). S. C. is supported by the Netherlands Organization for Scientific Research (NWO; VIDI project 639.042.017) and by the European Research Council (ERC; project PALs 320620).  P. C. and M. S. acknowledge the financial support of the European Research Council (ERC; project PALs 320620). We thank S. Hocuk and R. Garrod for stimulating discussions on dust temperature. We also thank the referee F. Le Petit for his helpful comments.

\end{acknowledgements}

{}


\begin{appendix}
\section{Tables}
\label{Tables}

We list in Tables \ref{table:adsorption_reactions}-\ref{table:cosmic_ray_reactions} all the surface reactions considered in the code and explained in Sect. \ref{Dust_chemistry}. In addition, we show the binding energies (Table \ref{table:binding_energies}) for each species depending on the type of the grain substrate (bare or icy). In Sect. \ref{Chemisorption}, we show the rates of the chemisorption reactions forming H$_2$.

\clearpage

\begin{table}
\caption{Adsorption reactions.}             
\centering 
\begin{tabular}{l l l}     
\hline\hline       
Reaction$^{(a)}$ &   & Reaction                           \\ 
\hline 
H $\rightarrow$   J(H)         & \vline & H$_2$O$_2$ $\rightarrow$ J(H$_2$O$_2$)            \\ 
H $\rightarrow$  J(H$_c$)      &\vline & CO $\rightarrow$  J(CO)         \\
H$_2$ $\rightarrow$  J(H$_2$)  &\vline & CO$_2$ $\rightarrow$ J(CO$_2$)             \\
O $\rightarrow$   J(O)         &\vline & HCO $\rightarrow$ J(HCO)         \\
O$_2$ $\rightarrow$  J(O$_2$)  &\vline & H$_2$CO $\rightarrow$ J(H$_2$CO)              \\
O$_3$ $\rightarrow$  J(O$_3$)  &\vline & H$_3$CO $\rightarrow$ J(H$_3$CO)              \\
OH $\rightarrow$  J(OH)        &\vline & CH$_3$OH $\rightarrow$ J(CH$_3$OH)           \\
H$_2$O $\rightarrow$ J(H$_2$O) &\vline &  N $\rightarrow$  J(N)         \\
HO$_2$ $\rightarrow$ J(HO$_2$) &  \vline  & N$_2$ $\rightarrow$  J(N$_2$) \\ 
\hline 
\hline 
\label{table:adsorption_reactions}                 
\end{tabular}
\tablefoot{$^{(a)}$ The expression J(i) means ice formation of the species $i$.\\
}
\end{table}

\begin{table}
\caption{Desorption reactions.}             
\centering 
\begin{tabular}{l l l }     
\hline\hline       
Reaction$^{(a)}$ & & Reaction  \\ 
\hline 
J(H)          $\rightarrow$  H      &  \vline  &  J(H$_2$O$_2$) $\rightarrow$  H$_2$O$_2$ \\
J(H$_c$)      $\rightarrow$  H      &  \vline  &  J(CO)         $\rightarrow$  CO  \\  
J(H$_2$)      $\rightarrow$  H$_2$  &  \vline  &  J(CO$_2$)     $\rightarrow$  CO$_2$   \\
J(O)          $\rightarrow$  O      &  \vline  &  J(HCO)        $\rightarrow$  HCO \\  
J(O$_2$)      $\rightarrow$  O$_2$  &  \vline  &  J(H$_2$CO)    $\rightarrow$  H$_2$CO\\
J(O$_3$)      $\rightarrow$  O$_3$  &  \vline  &  J(H$_3$CO)    $\rightarrow$  H$_3$CO\\ 
J(OH)         $\rightarrow$  OH     &  \vline  &  J(CH$_3$OH)    $\rightarrow$  CH$_3$OH\\
J(H$_2$O)     $\rightarrow$  H$_2$O &  \vline  &  J(N)      $\rightarrow$  N \\ 
J(HO$_2$)     $\rightarrow$  OH + O &  \vline & J(N$_2$)      $\rightarrow$  N$_2$\\
\hline 
\hline 
\label{table:desorption_reactions}                 
\end{tabular}
\tablefoot{$^{(a)}$ The expression J(i) means ice of the species $i$.\\
}
\end{table}

\begin{table}
\caption{Binding energies for the bare grain and water ice substrates.}             
\centering   
\begin{tabular}{l l l l }     
\hline\hline       
Species & $E$$_{\mathrm{b}}$ (K) & $E$$_{\mathrm{i}}$ (K) & References \\   
\hline                    
   H            & 500   & 650     &  (1), (2), (3)      \\
   H$_c$        & 10000 & 10000   &  (4)               \\
   H$_2$        & 300   & 300     &  (5)               \\
   O            & 1700  & 1700    &  (5)                \\
   O$_2$        & 1250  & 900     &  (3), (5), (6)        \\
   O$_3$        & 2100  & 1800    &  (3), (5)            \\
   OH           & 1360  & 3500    &  (3)                 \\
   H$_2$O       & 4800  & 4800    &  (9), (10)           \\
   HO$_2$       & 4000  & 4300    &  (3), (5)           \\
   H$_2$O$_2$   & 6000  & 5000    &  (3), (5)          \\ 
   CO           & 1100  & 1300    &  (2), (6), (7), (8)\\
   CO$_2$       & 2300  & 2300    &  (6)               \\
   HCO          & 830   & 3100    &  (10), (12)       \\
   H$_2$CO      & 1100  & 3100    &  (10), (12)\\
   H$_3$CO      & 1100  & 3100    &  (10), (12)\\
   CH$_3$OH     & 1100  & 3100    &  (11), (12)\\
   N            & 720   & 720     &  (13)                  \\  
   N$_2$        & 790   & 1140    &  (14), (15)             \\  
\hline
\hline 
\label{table:binding_energies}                  
\end{tabular}
\tablebib{
(1) Cazaux et al. (2010); (2) Garrod \& Pauly (2011); (3) Cuppen \& Herbst (2007); (4) Cazaux \& Tielens (2004); (5) Dulieu et al. (2013); (6) Noble et al. (2012a); (7) Collings et al. (2003); (8) Karssemeijer et al. (2014b); (9) Sandford \& Allamandola (1988); (10) Noble et al. (2012b); (11) we relate these to HCO$^{(5)}$ binding energies; (12) we relate these to CO$^{(2)(7)}$ binding energies; (13) Minissale et al. (2015); (14) Fuchs et al. (2006); (15) extrapolated from Kimmel (2001).}\\
\end{table}

\begin{table}
\caption{Reactions on grain surfaces.}             
\centering 
\begin{tabular}{l l l l}     
\hline\hline       
Reaction$^{(a)}$ & $\delta$$_{\mathrm{bare}}$$^{(b)}$ & $\delta$$_{\mathrm{ice}}$$^{(b)}$ & $\epsilon$$^{(c)}$          \\ 
\hline
J(H) + J(H)           $\rightarrow$  H$_2$              & 0.900 & 0.0900 & 0\\
J(H) + J(O)           $\rightarrow$  OH                 & 0.390 & 0.0390 & 0\\
J(H) + J(OH)          $\rightarrow$  H$_2$O             & 0.270 & 0.0270 & 0\\
J(H) + J(O$_3$)       $\rightarrow$  OH + O$_2$         & 0.080 & 0.0080 & 480\\
J(H) + J(H$_2$O$_2$)  $\rightarrow$  H$_2$O + OH        & 0.021 & 0.0021 & 1000\\
J(H) + J(CO)          $\rightarrow$  HCO                & 0.007 & 0.0007 & 5000\\
J(H) + J(HCO)         $\rightarrow$  H$_2$CO            & 0.070 & 0.0070 & 0\\
J(H) + J(HCO)         $\rightarrow$  CO + H$_2$         & 0.470 & 0.0470 & 0\\
J(H) + J(H$_3$CO)     $\rightarrow$  CH$_3$OH           & 0.023 & 0.0023 & 0 \\
J(O) + J(O)           $\rightarrow$  O$_2$              & 0.680 & 0.0680 & 0\\
J(N) + J(N)           $\rightarrow$  N$_2$              & 0.890 & 0.0890 & 0\\
\hline
J(H) + J(H)           $\rightarrow$  J(H$_2$)           & 0.100 & 0.9100 & 0\\
J(H) + J(O)           $\rightarrow$  J(OH)              & 0.610 & 0.9610 & 0\\
J(H) + J(OH)          $\rightarrow$  J(H$_2$O)          & 0.730 & 0.9730 & 0 \\
J(H) + J(O$_3$)       $\rightarrow$  J(OH) + J(O$_2$)   & 0.920 & 0.9920 & 480\\
J(H) + J(H$_2$O$_2$)  $\rightarrow$  J(H$_2$O) + J(OH)  & 0.979 & 0.9979 & 1000\\  
J(H) + J(CO)          $\rightarrow$  J(HCO)             & 0.993 & 0.9993 & 5000\\
J(H) + J(HCO)         $\rightarrow$  J(H$_2$CO)         & 0.930 & 0.9930 & 0\\
J(H) + J(H$_3$CO)     $\rightarrow$  J(CH$_3$OH)        & 0.977 & 0.9977 & 0\\
J(O) + J(O)           $\rightarrow$  J(O$_2$)           & 0.320 & 0.9320 & 0\\
J(N) + J(N)           $\rightarrow$  J(N$_2$)           & 0.110 & 0.9110 & 0\\
J(H) + J(HCO)         $\rightarrow$  J(CO) + J(H$_2$)   & 0.530 & 0.9530 & 130\\
J(H) + J(H$_2$CO)     $\rightarrow$  J(H$_3$CO)         & 1.000 & 1.0000 & 6000\\
J(H) + J(H$_2$CO)     $\rightarrow$  J(HCO) + J(H$_2$)  & 1.000 & 1.0000 & 2200\\
J(H) + J(H$_3$CO)     $\rightarrow$  J(H$_2$CO) + J(H$_2$)& 1.000 & 1.0000 & 150\\
J(H) + J(CH$_3$OH)     $\rightarrow$  J(H$_3$CO) + J(H$_2$)& 1.000 & 1.0000 & 3200\\
J(H) + J(CO$_2$)      $\rightarrow$  J(CO) + J(OH)        & 1.000 & 1.0000 & 10000\\
J(H) + J(O$_2$)       $\rightarrow$  J(HO$_2$)            & 1.000 & 1.0000 & 0\\
J(H) + J(H$_2$O)      $\rightarrow$  J(OH) + J(H$_2$)     & 1.000 & 1.0000 & 9600\\
J(H) + J(HO$_2$)      $\rightarrow$  J(OH) + J(OH)        & 1.000 & 1.0000 & 0\\
J(O) + J(CO)          $\rightarrow$  J(CO$_2$)            & 1.000 & 1.0000 & 650\\
J(O) + J(HCO)         $\rightarrow$  J(CO$_2$) + J(H)     & 1.000 & 1.0000 & 0\\
J(O) + J(H$_2$CO)     $\rightarrow$  J(CO$_2$) + J(H$_2$) & 1.000 & 1.0000 & 335\\
J(O) + J(O$_3$)       $\rightarrow$  J(O$_2$) + J(O$_2$)  & 1.000 & 1.0000 & 2500\\
J(O) + J(HO$_2$)      $\rightarrow$  J(O$_2$) + J(OH)     & 1.000 & 1.0000 & 0\\
J(O) + J(OH)          $\rightarrow$  J(O$_2$) + J(H)      & 1.000 & 1.0000 & 0\\
J(O) + J(O$_2$)       $\rightarrow$  J(O$_3$)             & 1.000 & 1.0000 & 0\\
J(OH) + J(CO)         $\rightarrow$  J(CO$_2$) + J(H)     & 1.000 & 1.0000 & 400\\
J(OH) + J(HCO)        $\rightarrow$  J(CO$_2$) + J(H$_2$) & 1.000 & 1.0000 & 0\\
J(OH) + J(OH)         $\rightarrow$  J(H$_2$O$_2$)        & 1.000 & 1.0000 & 0\\
J(OH) + J(H$_2$)      $\rightarrow$  J(H) + J(H$_2$O)     & 1.000 & 1.0000 & 2100\\
J(OH) + J(CH$_3$OH)    $\rightarrow$  J(H$_3$CO) + J(H$_2$O) & 1.000 & 1.0000 & 1000\\
J(HO$_2$) + J(H$_2$)  $\rightarrow$  J(H) + J(H$_2$O$_2$) & 1.000 & 1.0000 & 5000\\
\hline
\hline 
\label{table:two-body-reactions}                 
\end{tabular}
\tablefoot{$^{(a)}$ The expression J(i) means ice of the species $i$. The reactions above the horizontal line produce gas-phase species, while those below the horizontal line produce solid species.  $^{(b)}$ The parameters $\delta$$_{\mathrm{bare}}$ and $\delta$$_{\mathrm{ice}}$ indicate the probabilities of desorption upon reaction for bare and icy substrates, respectively. For each substrate, the sum of probabilities of two reactions giving the same species (gas-phase species from one reaction and solid species from the other one) is equal to 1. $^{(c)}$ The parameter $\epsilon$ indicates the activation barrier for each reaction. \\
}
\end{table}

\begin{table*}
\caption{Photoreactions on dust grains.}             
\centering   
\begin{tabular}{l l l}     
\hline\hline       
Reactions$^{(a)}$ & $\alpha$$_{{i}}$$^{(b)}$ (s$^{-1}$) & $\xi$$_{{i}}$$^{(b)}$ \\   
\hline                    
J(H$_2$)  + Photon     $\rightarrow$ J(H) + J(H)      & 3.40$\times$10$^{-11}$ & 2.50    \\
J(O$_2$)  + Photon     $\rightarrow$ J(O) + J(O)      & 7.90$\times$10$^{-10}$ & 2.13 \\
J(OH)     + Photon     $\rightarrow$ J(H) + J(O)      & 3.9$\times$10$^{-10}$ & 2.24  \\
J(CO$_2$) + Photon     $\rightarrow$ J(O) + J(CO)     & 8.9$\times$10$^{-10}$ & 3.00   \\
J(H$_2$O) + Photon     $\rightarrow$ J(H) + J(OH)     & 8.00$\times$10$^{-10}$ & 2.20   \\
J(HCO)    + Photon     $\rightarrow$ J(H) + J(CO)     & 1.10$\times$10$^{-09}$ & 1.09  \\
J(H$_2$CO)+ Photon     $\rightarrow$ J(H) + J(HCO)    & 5.87$\times$10$^{-10}$ & 0.53   \\
J(H$_3$CO)+ Photon     $\rightarrow$ J(H) + J(H$_2$CO) & 5.87$\times$10$^{-10}$ & 0.53  \\
J(H$_4$CO)+ Photon     $\rightarrow$ J(H) + J(H$_3$CO) & 5.87$\times$10$^{-10}$ & 0.53     \\
J(HO$_2$) + Photon     $\rightarrow$ J(O) + J(OH)      & 3.28$\times$10$^{-10}$  & 1.63 \\
J(HO$_2$) + Photon     $\rightarrow$ J(O$_2$) + J(H)   & 3.28$\times$10$^{-10}$  & 1.63 \\
J(H$_2$O$_2$) + Photon $\rightarrow$ J(OH) + J(OH)     & 8.30$\times$10$^{-10}$  & 1.80 \\
J(O$_3$)      + Photon $\rightarrow$ J(O$_2$) + J(O)   & 3.30$\times$10$^{-10}$  & 1.40 \\
J(N$_2$)  + Photon     $\rightarrow$ J(N) + J(N)       & 2.30$\times$10$^{-12}$  & 3.88 \\ 
\hline
J(H$_2$O) + Photon     $\rightarrow$ H$_2$O            & 2.16$\times$10$^{-11}$  & 2.20 \\
J(H$_2$CO)+ Photon     $\rightarrow$ H$_2$CO           & 2.16$\times$10$^{-11}$  & 0.53 \\
J(CO)     + Photon     $\rightarrow$ CO                & 2.20$\times$10$^{-15}$  & 2.54 \\
\hline 
\label{table:photo-reactions}                  
\end{tabular}
\tablefoot{$^{(a)}$ The expression J(i) means ice of the species $i$. $^{(b)}$ Values for $\alpha$$_{{i}}$ and $\xi$$_{{i}}$ (dimensionless) are taken from KIDA. 
\\
}
\end{table*}

\begin{table*}[h!]
\caption{Cosmic-ray reactions.}             
\centering 
\begin{tabular}{l l  l l l }     
\hline\hline       
Reaction$^{(a)}$  &  $\kappa$$_{\mathrm{CR}}$$^{(b)}$ (s$^{-1}$)& \vline  & Reaction$^{(a)}$ &    $\kappa$$_{\mathrm{CR}}$$^{(b)}$ (s$^{-1}$)                \\ 
\hline 
J(H$_2$)  + CR       $\rightarrow$  J(H) + J(H)         &  5.00$\times$10$^{-17}$   & \vline &  J(CO)     + CRP    $\rightarrow$  CO              & 1.08$\times$10$^{-14}$ \\
J(O$_2$)  + CRP      $\rightarrow$  J(O)  + J(O)        &  3.75$\times$10$^{-14}$   & \vline &  J(H$_2$O) + CRP    $\rightarrow$  H$_2$O          & 1.08$\times$10$^{-14}$\\
J(OH)     + CRP      $\rightarrow$  J(H) + J(O)         &  2.55$\times$10$^{-14}$   & \vline &  J(H$_2$CO)+ CRP    $\rightarrow$  H$_2$CO         & 1.08$\times$10$^{-14}$ \\
J(CO$_2$) + CRP      $\rightarrow$  J(O) + J(CO)        &  8.55$\times$10$^{-14}$   & \vline &  J(CH$_3$OH)+ CRP    $\rightarrow$  CH$_3$OH        & 1.08$\times$10$^{-14}$  \\
J(H$_2$O) + CRP      $\rightarrow$  J(H) + J(OH)        &  4.85$\times$10$^{-14}$   & \vline &  J(N$_2$)  + CRP    $\rightarrow$  J(N) + J(N)     & 2.50$\times$10$^{-16}$  \\ 
J(HCO)    + CRP      $\rightarrow$  J(H) + J(CO)        &  2.11$\times$10$^{-14}$   & \vline &  J(HO$_2$) + CRP    $\rightarrow$  J(O) + J(OH)    & 3.75$\times$10$^{-14}$ \\
J(H$_2$CO)+ CRP      $\rightarrow$  J(H) + J(HCO)       &  2.11$\times$10$^{-14}$   & \vline &  J(HO$_2$) + CRP    $\rightarrow$  J(H) + J(O$_2$) & 3.75$\times$10$^{-14}$  \\
J(H$_3$CO)+ CRP      $\rightarrow$  J(H) + J(H$_2$CO)   &  2.11$\times$10$^{-14}$   & \vline &  J(H$_2$O$_2$) + CRP $\rightarrow$ J(OH) + J(OH)   & 7.50$\times$10$^{-14}$ \\
J(CH$_3$OH)+ CRP      $\rightarrow$  J(H) + J(H$_3$CO)   &  2.11$\times$10$^{-14}$   & \vline &  J(O$_3$)  + CRP    $\rightarrow$  J(O$_2$) + J(O) & 3.75$\times$10$^{-14}$ \\
\hline 
\hline 
\label{table:cosmic_ray_reactions}                 
\end{tabular}
\tablefoot{$^{(a)}$ The expression J(i) means ice of the species $i$. $^{(b)}$ Values for the cosmic ray rate coefficient, $\kappa$$_{\mathrm{CR}}$, are taken from KIDA.\\
}
\end{table*}

\clearpage

\subsubsection{Chemisorption reactions}
\label{Chemisorption}

We also consider in the code the chemical reaction JH$\rightarrow$JH$_{\mathrm{c}}$ and the following reactions forming H$_2$ from chemisorbed hydrogen (H$_{\mathrm{c}}$): 
 

\begin{equation}
\mathrm{H + J(H_c) \rightarrow H_2} 
\label{equation:H2_6302}
\end{equation}

\noindent and

\begin{equation}
\mathrm{J(H) + J(H_c) \rightarrow H_2}, 
\label{equation:H2_6305}
\end{equation}

\noindent where J(X) means solid X. 
The transmission coefficient through a barrier width $a$, with energy $\epsilon$(K), is:

\begin{equation}
P_{\mathrm{chem}} = exp\left(-a\displaystyle{ \sqrt {\frac{2m_{red}k_{\mathrm{B}}\epsilon}{\hbar^{2}}}}\right) + exp\left(\displaystyle{ \frac{-\epsilon}{T_{\mathrm{g}}}}\right), 
\end{equation}

\noindent where $\hbar$ is the Planck constant divided by 2$\pi$, $m$$_{\mathrm{red}}$ is the reduced mass of the reactants, and $k$$_{\mathrm{B}}$ is the Boltzmann constant. The chemisorption reaction rate, $R$$_{\mathrm{chem}}$ (cm$^{-3}$s$^{-1}$), for (\ref{equation:H2_6302}) is given by

\begin{equation}
R_{\mathrm{chem}}= k_{\mathrm{ads}} P_{\mathrm{chem}} f_{\mathrm{bare}} S(T_{\mathrm{g}}, T_{\mathrm{d}}) \displaystyle{ \sqrt {\frac{T_{\mathrm{g}}}{100}}}, 
\end{equation}

\noindent where $k$$_{\mathrm{ads}}$ is the adsorption rate coefficient (see Sect. \ref{Adsorption}) and $S$($T$$_{\mathrm{g}}$, $T$$_{\mathrm{d}}$) the sticking coefficient. 

When the formation of H$_2$ is through a chemical reaction between solid physisorbed and chemisorbed hydrogen (reaction \ref{equation:H2_6305}), it is necessary to consider the binding energies of H and H$_{\mathrm{c}}$ for bare grains ($E$$_{\mathrm{bare}}$). In this case, the chemisorption reaction rate, $R$$_{\mathrm{chem}}$ (cm$^{-3}$s$^{-1}$) is given by

\begin{equation}
R_{\mathrm{chem}}= \nu_0 \left(\displaystyle{ \frac{n_{\mathrm{i}}n_{\mathrm{j}}}{n_{\mathrm{d}}n_{\mathrm{sites}}}}\right) f_{\mathrm{bare}} k_{\mathrm{chem}}. 
\end{equation}

\noindent $\nu$$_0$ is the oscillation factor and $k$$_{\mathrm{chem}}$ determines the probability of mobility by tunneling ($m$$_{\mathrm{tunn}}$) and thermal diffusion ($m$$_{\mathrm{diff}}$)

\begin{equation}
k_{\mathrm{chem}}= m_{\mathrm{tunn}} + m_{\mathrm{tunn}} = [8 \sqrt{\pi T_{\mathrm{d}}} (A \times B)] + [4C],  
\end{equation}

\noindent with

\begin{equation}
A= \displaystyle{ \frac{\sqrt{E_{\mathrm{bare}}(J(H_{\mathrm{c}}))-E_{\mathrm{bare}}(J(H))}}{E_{\mathrm{bare}}(J(H_{\mathrm{c}}))+\epsilon}}, 
\end{equation}

\begin{equation}
B= exp\left(-0.406 \times d \times \displaystyle{ \sqrt{E_{\mathrm{bare}}(J(H_{\mathrm{c}}))+\epsilon}} \right), 
\end{equation}

\noindent where $d$ is the distance between physisorbed and chemisorbed sites (we assume 3$\AA$), and
 
\begin{equation}
C= \displaystyle{ \sqrt {\frac{E_{\mathrm{bare}}(J(H)) + \epsilon}{E_{\mathrm{bare}}(J(H_{\mathrm{c}}))+\epsilon}}} \times exp\left(\displaystyle{ \frac{-[E_{\mathrm{bare}}(J(H)) + \epsilon]}{T_{\mathrm{d}}}}\right). 
\end{equation}

\noindent See Cazaux $\&$ Tielens (2004) for more details about these rates.

\clearpage

\end{appendix}

\end{document}